\documentclass{jfm}

\usepackage{graphicx}
\usepackage{newtxtext}
\usepackage{newtxmath}
\usepackage{natbib}
\usepackage{hyperref}
\hypersetup{
    colorlinks = true,
    urlcolor   = blue,
    citecolor  = black,
}

\newcommand{\RomanNumeralCaps}[1]
\linenumbers
\usepackage{comment}
\usepackage{color}
\usepackage{subcaption}
\usepackage{bm}

\usepackage[OMLmathsfit,sfdefault=cmbr]{isomath}
\newcommand\mathtensor[1]{\mathsfbfit{#1}}

\newcommand{\Wi}{\textit{Wi}}
\newcommand{\Sc}{\textit{Sc}}

\shorttitle{Structures of elastoinertial turbulence in pipe flow}
\shortauthor{M. Kumar and M. D. Graham}

\title{Structures of elastoinertial turbulence in pipe flow}

\author{Manish Kumar\aff{1} \corresp{\email{manish030297@gmail.com}} \and Michael D. Graham\aff{1}}

\affiliation{\aff{1}Department of Chemical and Biological Engineering, University of Wisconsin-Madison, 1415 Engineering Dr, Madison, WI 53706, USA
\\[\affilskip]
}

\begin{document}

\maketitle
\begin{abstract}
Elastoinertial turbulence (EIT) is a self-sustaining chaotic state resulting from the interplay between inertia and elasticity in the flow of dilute polymeric solutions, and its emergence is believed to limit the achievable drag reduction in turbulence flow using polymer additives. In the present study, we introduce a viscoelastic variant of spectral proper orthogonal decomposition (VESPOD) that decomposes velocity and polymeric stress fields of EIT together into well-defined orthogonal oscillating modes such that the decomposition is optimal in the terms of the total mechanical energy of the flow. Using this technique, we investigate the dominant coherently evolving structures underlying the dynamics of EIT in axisymmetric pipe flow. By analyzing distinct peaks in the leading eigenvalue of the VESPOD eigenvalue spectrum, we find that the dynamics of EIT in pipe flow is dominated by three distinct families of traveling waves, where the higher wavenumber structures of each family are simple harmonics of their respective fundamental waves. The radial velocity fields of the traveling waves are characterized by the formation of large-scale structures spanning the pipe radial direction. However, the polymeric stress fields corresponding to them are characterized by the formation of thin inclined sheets of high stress fluctuations at the critical layers of the respective waves, i.e.~ the locations where the wave speed of the VESPOD mode matches the mean streamwise velocity. Additionally, these sheets exhibit nested structures, where the polymeric sheets of faster waves are confined by those of the immediately slower waves.       

\end{abstract}

\section{Introduction}
The reduction of turbulent drag using polymer additives has been an active area of research due to its application to save energy during liquid transport \citep{Xi:2019kt} ever since it was discovered that the addition of a tiny amount ($\sim O(10)$ ppm) of high molecular weight polymer suppresses inertial turbulence and dramatically reduces frictional drag \citep{toms1949some}. Although polymer additives can significantly reduce frictional drag, the flow does not become laminar; hence, the drag is always higher than the laminar value. This phenomenon is attributed to the presence of a self-sustaining weakly fluctuating flow state known as the Maximum Drag Reduction (MDR) asymptote \citep{Virk:1970uo}. At MDR, the frictional drag is nearly independent of the type and concentration of polymers; understanding MDR is an important area of research because it sets a limit for attainable amount of  drag reduction. Earlier studies of MDR mainly focused on the ensemble statistics of the velocity fluctuations, however, recent advancements in the field attempt to understand the dynamic evolution of the MDR state \citep{Xi:2019kt}. The discovery of elastoinertial turbulence (EIT), a self-sustaining chaotic state resulting from the interplay between inertia and elasticity, provides some understanding of the dynamics of the MDR state \citep{Samanta2013,Datta2022,Dubief2023}. In this paper, we investigate the dynamics of EIT in pipe flow and discover dominant structures underlying its dynamics.

Elastoinertial turbulence can sustain in the parameter regime characterized by the Reynolds number $\Rey$ (dimensionless number representing the ratio of inertial to viscous forces) and the Weissenberg number $\Wi$ (product of polymer relaxation time to characteristic strain rate), where the flow is linearly stable, suggesting a nonlinear route for transition to EIT. Viscoelastic \textcolor{black}{two-dimensional (2-D)} channel flow becomes linearly unstable due to a wall mode instability at sufficiently large $\Rey$ \citep{Zhang2013} and a center mode instability at a large $\Wi$ \citep{Page2020, Khalid2021}, the parameter values significantly higher than the EIT parameter regime of interest here. The wall mode instability leads to the emergence of a traveling wave known as the ``Tollmien–Schlichting (TS)" wave \citep{drazin1981hydrodynamic}, and the center mode instability leads to the emergence of another distinct traveling wave having an ``arrowhead" structure \citep{Page2020}. The origins of these traveling waves are highly subcritical, and they can persist subcritically to the parameter regimes relevant to EIT. Inspired by these facts, it was speculated that the non-linear excitation of either the wall mode \citep{Shekar2019,Shekar2020,Shekar2021} or the center mode \citep{Page2020,Dubief2022} could lead to EIT in 2-D. Recent advancements indicate that in elastoinertial regime, wall-mode structures dominate the dynamics \citep{Kumar2024,Kumar2025,Beneitez2024}, while at vanishing $\Rey$, the dynamics tend to be organized around the center mode structures \citep{Morozov2022,Beneitez2024PRF,Lewy2025}.

Recently, using spectral proper orthogonal decomposition (SPOD) \citep{Towne2018}, we have revealed the dominant coherent structures underlying the dynamics of EIT in 2-D channel flow \citep{Kumar2024}. The SPOD is a temporal variant of the proper orthogonal decomposition (POD) \citep{Holmes_Lumley_Berkooz_Rowley_2012} and decomposes time-dependent state variables into oscillating orthogonal basis functions such that the basis functions are coherent in both space and time. We demonstrated that the dynamics of EIT in channel flow is predominantly composed of a family of non-harmonic traveling waves, where the most dominant traveling wave of the family resembles the structure of the Tollmien–Schlichting (TS) wave \citep{Kumar2024}. The wall-normal velocity of the most dominant traveling wave has large-scale regular structures spanning the channel height, the streamwise velocity component has regular structures in the vicinity of the channel walls, and the polymeric stress field contains thin sheets of high polymeric stress fluctuations localized at the critical layers of the traveling wave. The critical layer is the location where the wave speed of the traveling wave matches the mean streamwise velocity. The remaining traveling waves of the family exhibit similar structures, however nested in the region bounded by the polymeric sheets and hence critical layers of immediately slower traveling wave. The sheets of high polymeric stress act as barriers for flow and hence prevent velocity fluctuations from penetrating them \citep{Kumar2023stretching}. Thus, the dynamics of EIT in channel flow originates from a nonlinear excitation of a wall mode and is dominated by a family of self-similar nested traveling waves where the polymeric sheets of each traveling wave confines the velocity fluctuations of their immediate faster wave. Strengthening the notion of the connection of EIT with wall mode, it has also been found that the arrowhead structure resulting from the center mode instability is disconnected from EIT, and it does not play a role in the self-sustenance of the dynamics of EIT \citep{Beneitez2024}.

The \textcolor{black}{numerical} investigation of the dynamics of EIT in pipe flow has received relatively little attention \citep{Lopez:2019ct} and is the topic of the present study. Similar to channel flow, the center mode becomes linearly unstable even in viscoelastic pipe flow at $\Rey=O(100)$ and $\Wi=O(10)$ \citep{Garg2018,Chaudhary2021}. Weakly nonlinear analysis around the linearly unstable center-mode demonstrates that nonlinearity stabilizes the flow at high polymer concentration while it destabilizes the flow at low polymer concentration, indicating supercritical and subcritical bifurcations at high and low polymer concentrations, respectively \citep{Wan2021}.  In contrast, the wall mode in the pipe flow remains linearly stable at all $\Rey$. However, using resolvent analysis, it has been found that the structure of the most strongly amplified mode \textcolor{black}{by the linearized dynamics} in viscoelastic pipe flow contains strong stress fluctuations localized at the critical layer \citep{Zhang2021}, similar to the viscoelastic channel flow \citep{Shekar2019}, indicating a critical layer mechanism of the dynamics of EIT even in pipe flow. Further, there are common characteristics of the structures of EIT in both channel and pipe flows. For example, EIT in both channel and pipe flows exhibits near-wall vortical structures oriented in the direction perpendicular to the mean flow direction and near-wall elongated streaks of streamwise velocity fluctuations aligned in the flow direction with a slight slope \citep{Samanta2013,Lopez:2019ct,Choueiri2021}. 


The basic characteristics of EIT in channel and pipe flows are fundamentally 2-D \citep{Sid2018,Lopez:2019ct}. Therefore, in the present study, we investigate the dynamics of EIT in axisymmetric pipe flow. 
We perform direct numerical simulations of EIT using a spectral method. To analyze the dynamics, we introduce a viscoelastic variant of the SPOD \citep{Towne2018}, which we refer to as ``VESPOD", that enables us to investigate the well-defined structures underlying EIT that evolve coherently in time. As detailed below, VESPOD is based on the total mechanical energy (kinetic plus elastic) of the flow, enabling an energy-based modal decomposition that contains both velocity and polymer stress fields.  The eigenvalue spectrum of VESPOD exhibits distinct peaks and by analyzing those peaks we find that the dynamics of EIT in pipe flow, in the parameter regime studied, is dominated by three distinct families of traveling waves. We also explore the effects of the domain length, $\Rey$, and $\Wi$ on the dynamics of EIT and the traveling waves underlying its dynamics and comment on the relation between EIT in pipe and channel flow.  The remaining part of the paper has been organized as follow: Section \ref{formulation} discusses the methodology, Section \ref{results} focuses on results and discussion, and Section \ref{conclusion} concludes the finding of the study.


\section{Formulation and governing equations}\label{formulation}
\subsection{Direct numerical simulation of EIT}\label{DNS}
The flow is governed by the conservation of momentum and mass, described by the equations: 
\begin{equation}\label{colm}
\frac{\partial \boldsymbol{u}}{\partial t}+\boldsymbol{u}\cdot \nabla \boldsymbol{u} =-\nabla p+\frac{\beta}\Rey \nabla^2 \boldsymbol{u} + \frac{1-\beta}\Rey \nabla \cdot \boldsymbol{\tau}_p+F(t)\boldsymbol{e}_z,\qquad \nabla \cdot \boldsymbol{u}=0, 
\end{equation}
where $\boldsymbol{u}$ and $p$ represent non-dimensional velocity field and pressure field, respectively. The Reynolds number is defined as $\Rey=\rho U_c R/\eta$, where $\rho$ and $\eta$ are the density and zero-shear rate viscosity of the fluid. The pipe radius ($R$) and the Newtonian laminar centerline velocity ($U_c$) are the characteristic length and velocity scales, respectively. The viscosity ratio $\beta$  represents the contribution of solvent viscosity ($\eta_s$) to the zero-shear solution viscosity and is defined as $\beta=\eta_s/\eta$. The contribution of polymer chains to the stress tensor in the momentum equation is denoted as $\boldsymbol{\tau}_p$, and we use the FENE-P constitutive equation having polymer molecular diffusion to model it \citep{bird1987dynamics}: 
\begin{equation}\label{tau_theta}
\frac{\partial \boldsymbol{\alpha}}{\partial t}+\boldsymbol{u}\cdot \nabla \boldsymbol{\alpha} - \boldsymbol{\alpha}\cdot \nabla \boldsymbol{u}-(\boldsymbol{\alpha}\cdot \nabla \boldsymbol{u})^T=-\boldsymbol{\tau}_p+\frac{1}{\Rey \Sc} \nabla^2 \boldsymbol{\alpha},
\end{equation}
\begin{equation}\label{fenep}
\boldsymbol{\tau}_p= \frac{1}{\Wi} \left( \frac{\boldsymbol{\alpha}}{1-\mathrm{tr}(\boldsymbol{\alpha})/b}-\mathsfbi{I}\right),
\end{equation}
where $\boldsymbol{\alpha}$, $\mathsfbi{I}$, and $b$ are the conformation tensor, the identity tensor, and the maximum extensibility of the polymer chain, respectively. The Weissenberg number is defined as $\Wi=\lambda U_c/R$, where $\lambda$ is the polymer relaxation time. We keep the molecular diffusion in the evolution equation of the conformation tensor (Eq. \ref{tau_theta}) to ensure numerical stability. The strength of molecular diffusion is controlled by the Schmidt number $Sc=\eta/\rho D$, where $D$ is the diffusion coefficient. 

We perform numerical simulations for the axisymmetric pipe flow EIT. We consider non-slip boundary conditions on the wall of the pipe ($r=1$) and axisymmetric boundary conditions at the center of the pipe ($r=0$). The finite molecular diffusion of the polymer chains requires boundary conditions for the governing equation of the conformation tensor (Eq. \ref{tau_theta}). At the wall ($r=1$), we solve Eq. \ref{tau_theta} considering $1/Sc=0$ and use the respective value of $\boldsymbol{\alpha}$ as the boundary condition. In the streamwise direction, we consider periodic boundary conditions. The volumetric flow through the pipe is kept constant at the Newtonian laminar value by tuning the body force ($F(t)$) at each time step.   

We used \emph{Dedalus} \citep{Burns2020} to perform direct numerical simulations (DNS) using the spectral method.  The code validation can be found in the Appendix \ref{code_validation}. We consider viscosity ratio $\beta=0.97$ representing extreme dilute limit and $b=6400$ corresponding to high molecular-weight polymer chains ($500$ kDa polyacrylamide). The length of computational domain is $L=5 \times n$, where $n$ is an integer. We used $N_z \times n = 256 \times n$ Fourier basis functions in the streamwise direction and $N_r = 512$ Chebyshev basis functions in the radial direction, respectively. \textcolor{black}{For time-stepping, we use a 2nd-order 2-stage implicit-explicit Runge-Kutta scheme  with a fixed time-step size $\Delta t =0.001$ (unless mentioned otherwise).} We consider molecular diffusion ($Sc=250$) similar to the previous studies \citep{Sid2018,Kumar2024}. The simulations have been initiated with sufficiently large random perturbations in the state variables to trigger EIT. The simulation data corresponding to the initial 200 time units are dropped to make sure that the dynamics are fully developed. \textcolor{black}{We also performed a sensitivity analysis of the EIT dynamics on $Sc$ and mesh resolution (Appendix \ref{mesh_resolution_study}), and continue with the parameters mentioned above as they capture the essential physics with affordable computational cost.}     

\subsection{VESPOD}\label{VESPOD}
To analyze the EIT dataset for pipe flow, we use a viscoelastic variant of SPOD (VESPOD), which allows the modal decomposition of the velocity and polymeric stress fields together. We note that the flow displays translation symmetry in the streamwise ($z$) direction, in which case the Fourier basis functions are optimal in terms of energy \citep{Holmes_Lumley_Berkooz_Rowley_2012}. To embed this symmetry in our VESPOD framework, first, we Fourier-transformed state variables in the streamwise direction and then perform VESPOD decomposition wavenumber-by-wavenumber. The VESPOD represents the frequency-by-frequency VEPOD decomposition of a Fourier-transformed (in temporal direction) time-dependent dataset \citep{Holmes_Lumley_Berkooz_Rowley_2012,Wang2014}, where VEPOD is a viscoelastic variant of POD \citep{Kumar2025}. Hence, VESPOD seeks to find the basis function $\boldsymbol{\psi}(\kappa,r,f)$ that maximizes the objective function 
\begin{equation}\label{objective_fn}
E\{ | \langle \boldsymbol{q}(\kappa,r,f), \boldsymbol{\psi}(\kappa,r,f) \rangle |^2\}
\end{equation}
given the constraint $\langle \boldsymbol{\psi}(\kappa,r,f), \boldsymbol{\psi}(\kappa,r,f) \rangle=1$, where $E\{\cdot\}$ and $|\cdot|$ represent expectation and modulus operations on the data ensemble, respectively. The quantity $\boldsymbol{q}(\kappa,r,f)$ represents a vector containing state variables having \textcolor{black}{scaled} streamwise wavenumber \textcolor{black}{$\kappa$ (defined as the number of wavelengths per domain length)} and temporal oscillation frequency $f$. In VESPOD,  $\boldsymbol{q}$ is defined as
\begin{equation}\label{q_entry}
\boldsymbol{q}=\left[\boldsymbol{u}', \mathsfbi{T}'\right],
\end{equation}
where the stretch tensor $\mathsfbi{T}=\sqrt{\frac{1-\beta}{\mathit{Re} \Wi}}\boldsymbol{\theta}$ and $\boldsymbol{\theta} \cdot \boldsymbol{\theta}=\boldsymbol{\alpha}/(1-\mathrm{tr}(\boldsymbol{\alpha})/b)$ \citep{Wang2014}. The superscript ($'$) represents the perturbation from the mean (averaged over $t$ and $z$). The inner product $\langle \cdot , \cdot\rangle$ in the objective function \eqref{objective_fn} has been defined as: 
\begin{equation}\label{inner_product}
\langle\boldsymbol{q,\psi}\rangle= 2\pi\int_{0}^{1} \boldsymbol{q}(r) \cdot \boldsymbol{\psi}(r) r dr. 
\end{equation}
The specific definition of variable $\boldsymbol{q}$ leads to the following inner product of $\boldsymbol{q}$ with itself:  
\begin{equation}\label{mechanical_energy}
\langle\boldsymbol{q,q}\rangle= 2\pi\int_{0}^1 \left\{\boldsymbol{u}' \cdot \boldsymbol{u}' + \frac{1-\beta}{\Rey \Wi}\boldsymbol{\theta}' : \boldsymbol{\theta}'\right\}  rdr. 
\end{equation} 
This inner product and choice of $\boldsymbol{q}$ are chosen because in the limit $b\rightarrow\infty$, Eq.~\ref{mechanical_energy} becomes the total mechanical energy of the flow and is thus a natural and interpretable inner product for analysis of viscoelastic flows. In this limit, VESPOD yields an optimal decomposition in terms of total mechanical energy. For $b$ finite, the identification with total mechanical energy is approximate but is nevertheless useful. \textcolor{black}{We also plot the Probability Distribution Function (PDF) of $\mathrm{tr}(\boldsymbol{\alpha})/b$ to find the distribution of stretching of polymeric chains in EIT (Appendix \ref{pdf_polymeric_stress}). The probability of a large value of $\mathrm{tr}(\boldsymbol{\alpha})/b$ is small, indicating the inner product used in the present study provides a good representative of the total mechanical energy.}

The maximization of the objective function \eqref{objective_fn} leads to the following self-adjoint eigenvalue problem:
\begin{equation}\label{eigen_continous}
2\pi \int_{0}^1 E\{\boldsymbol{q}(\kappa,r,f) \boldsymbol{q}^*(\kappa,\rho,f) \}  \boldsymbol{\psi}(\kappa,\rho,f) \rho d \rho=\sigma  \boldsymbol{\psi}(\kappa,r,f),
\end{equation}
which gives an infinite set of eigenmodes $\{\sigma_j, \boldsymbol{\psi}_j(\kappa,r,f)\}$ which are arranged in decreasing value of (approximate) mechanical energy. The state variables can be reconstructed using VESPOD modes as 
\begin{equation}\label{reconstruct_POD} \Tilde{\boldsymbol{q}}(\kappa, r, f)=\sum_{j=1}^{\infty}a_j(\kappa,f)\boldsymbol{\psi}_j(\kappa, r, f),
\end{equation}
where $a_j(\kappa,f)=\langle \boldsymbol{q}(\kappa, r, f), \boldsymbol{\psi}_j(\kappa, r, f) \rangle$.

To obtain the VESPOD of a discrete time series, the data matrix containing the Fourier coefficients of the structures having wavenumber $\kappa$ of the discrete time series of snapshots $N_t$ is constructed as:
\begin{equation}\label{matrix_Qk}
\mathtensor{Q}_{\kappa}=[\boldsymbol{q}_{1,\kappa}, \boldsymbol{q}_{2,\kappa}, ...., \boldsymbol{q}_{N_t, \kappa}],
\end{equation}
where $\boldsymbol{q}_{i,\kappa}=\boldsymbol{q}(\kappa,r,t_i)$. \textcolor{black}{The computation of VESPOD using Eq. \ref{eigen_continous} requires ensemble of independent realizations. In practice when only a single time resolved data is available, however the flow is chaotic (adequate temporally decorrelated), the ensemble can be constructed by using data from different time windows of a very long time series \citep{Welch1967}.} Therefore, multiple realizations of the flow time series have been generated by dividing the data matrix into overlapping blocks as
\begin{equation}\label{matrix_Qn}
\mathtensor{Q}^n_{\kappa}=[\boldsymbol{q}_{1,\kappa}^n, \boldsymbol{q}_{2,\kappa}^n, ..,\boldsymbol{q}_{m,\kappa}^n,.., \boldsymbol{q}_{N_f,\kappa}^n], \ n=1,2,..., N_b,
\end{equation}
where $N_b$ and $N_f$ are the number of blocks and the number of snapshots in each block. For $N_o$ overlapping snapshots in each block, the entries of the data matrix $\mathtensor{Q}^n_{\kappa}$ and $\mathtensor{Q}_{\kappa}$ are related as $\boldsymbol{q}_{m,\kappa}^n=\boldsymbol{q}_{m+(n-1)(N_f-N_o), \kappa}$. Next, we compute the discrete Fourier transform (DFT) of each block in the temporal direction. To minimize the spectral leakage resulting from the non-periodicity of the data in each block during the estimation of the DFT, we compute the DFT of the windowed data:    
\begin{equation}\label{matrix_Qnw}
\mathtensor{Q}^{n,w}_{\kappa}=[w_1\boldsymbol{q}_{1,\kappa}^n, w_2\boldsymbol{q}_{2,\kappa}^n, ..,w_m\boldsymbol{q}_{m,\kappa}^n,.., w_{N_f}\boldsymbol{q}_{N_f,\kappa}^n],
\end{equation}
where $w_m$ represents the nodal value of the symmetric Hamming window function and is given as:
\begin{equation}\label{hamming}
w_m= 0.54-0.46\cos \left(\frac{2\pi (m-1)}{N_f-1} \right).
\end{equation}
The Fourier coefficients of the temporal DFT of $\mathtensor{Q}^{n,w}_{\kappa}$ are given as
\begin{equation}\label{matrix_Qn_f}
\Tilde{\mathtensor{Q}}^n_{\kappa}=[\Tilde{\boldsymbol{q}}_{1,\kappa}^n, \Tilde{\boldsymbol{q}}_{2,\kappa}^n, ..,\Tilde{\boldsymbol{q}}_{m,\kappa}^n,.., \Tilde{\boldsymbol{q}}_{N_f,\kappa}^n],
\end{equation}
where $\Tilde{\boldsymbol{q}}_{m,\kappa}^n$ represents the Fourier coefficient of wavenumber $\kappa$ at frequency $f_m$ in the $n^{th}$ block. Now, the Fourier coefficients at the frequency $f_m$ are collected from each block and a new data matrix $\Tilde{\mathtensor{Q}}_{m,\kappa}$ is constructed as:
\begin{equation}\label{matrix_Qm_f}
\Tilde{\mathtensor{Q}}_{m,\kappa}=[\Tilde{\boldsymbol{q}}_{m,\kappa}^1, \Tilde{\boldsymbol{q}}_{m,\kappa}^2, .., \Tilde{\boldsymbol{q}}_{m,\kappa}^n, .., \Tilde{\boldsymbol{q}}_{m,\kappa}^{N_b}].
\end{equation}
The VESPOD modes at the frequency $f_m$ can be obtained by calculating the eigenmodes of the discretized cross-spectral density (CSD) matrix $\mathtensor{S}_{m\kappa}=\Tilde{\mathtensor{Q}}_{m,\kappa}\Tilde{\mathtensor{Q}}_{m,\kappa}^*$ by solving the following eigenvalue problem:
\begin{equation}\label{eigenvalue_discrete}
 \mathtensor{S}_{m,\kappa} \mathtensor{W} \boldsymbol{\psi}_{m,\kappa} = \boldsymbol{\psi}_{m,\kappa} \mathtensor{\sigma}_{m,\kappa},
\end{equation}
where $\boldsymbol{\psi}_{m,\kappa}=\boldsymbol{\psi}(\kappa,r,f_m)$ represents the VESPOD mode structures at wavenumber $\kappa$ and frequency $f_m$, and diagonal matrix $\mathtensor{\sigma}_{m,\kappa}$ represents mechanical energy associated with respective mode structures. The matrix $\mathtensor{W}$ is a positive-definite weighting matrix, which properly accounts for the integration on a non-uniform discrete grid.

The eigenmode computation of \eqref{eigenvalue_discrete} is expensive. To overcome this challenge, the following eigenvalue problem is solved:
\begin{equation}\label{eigenvalue_analogous}
\Tilde{\mathtensor{Q}}_{m,\kappa}^* \mathtensor{W} \Tilde{\mathtensor{Q}}_{m,\kappa} \boldsymbol{\Theta}_{m,\kappa} = \boldsymbol{\Theta}_{m,\kappa} \mathtensor{\sigma}_{m,\kappa}.
\end{equation}
This is faster to compute because the number of flow realizations ($N_b$) is much smaller than the number of grid points. This analogous problem gives the same nonzero eigenvalues as \eqref{eigenvalue_discrete}, and their eigenvectors are related as: 
\begin{equation}\label{eigenvector}
\boldsymbol{\psi}_{m,\kappa} = \Tilde{\mathtensor{Q}}_{m,\kappa}\boldsymbol{\Theta}_{m,\kappa} \mathtensor{\sigma}_{m,\kappa}^{-1/2}.
\end{equation}
The VESPOD mode structures having wavenumber $\kappa$ and frequency $f_m$ can be projected in the physical space as:
\begin{equation}\label{VESPOD_mode_rx}
\boldsymbol{\psi}_m(z,r) = \boldsymbol{\psi}_{m,\kappa} e^{i2\pi \kappa z/L}.
\end{equation}

We modify the MATLAB tool originally developed by \cite{Schmidt2022} to perform VESPOD analysis of EIT in pipe flow. \textcolor{black}{We used data corresponding to $600$ time units sampled at a time interval of $\Delta t_s=0.025$ time units, leading to a total of $24000$ snapshots. The multiple realizations of the flow \textcolor{black}{for the most of the cases} have been constructed by dividing the data into blocks having $N_f=2000$ snapshots in each block with $50\%$ overlap. This leads to a total of $23$ blocks, which is sufficient for a converged computation of VESPOD. The current choice of $\Delta t_s$ and $N_f$ leads to frequency resolution $\Delta f=1/(\Delta t_s N_f)=0.02$. However, for $\kappa=1$ structures, we have used $\Delta f=0.01$ by considering $N_f=4000$ to properly resolve distinct peaks in the eigenvalue spectra. A coarse frequency resolution may fail to resolve distinct spectral peaks, whereas an excessively fine frequency resolution can reduce statistical convergence and can generate numerous closely spaced peaks by resolving weak temporal variability within broadband structures, thereby making it difficult to isolate the dominant energetic structures. Therefore, an optimal value of $\Delta f$ is chosen to balance frequency resolution and statistical convergence (Appendix \ref{VESPOD_spectra_different_nfft}).}

\begin{figure}
\centering
\includegraphics[width=\textwidth]{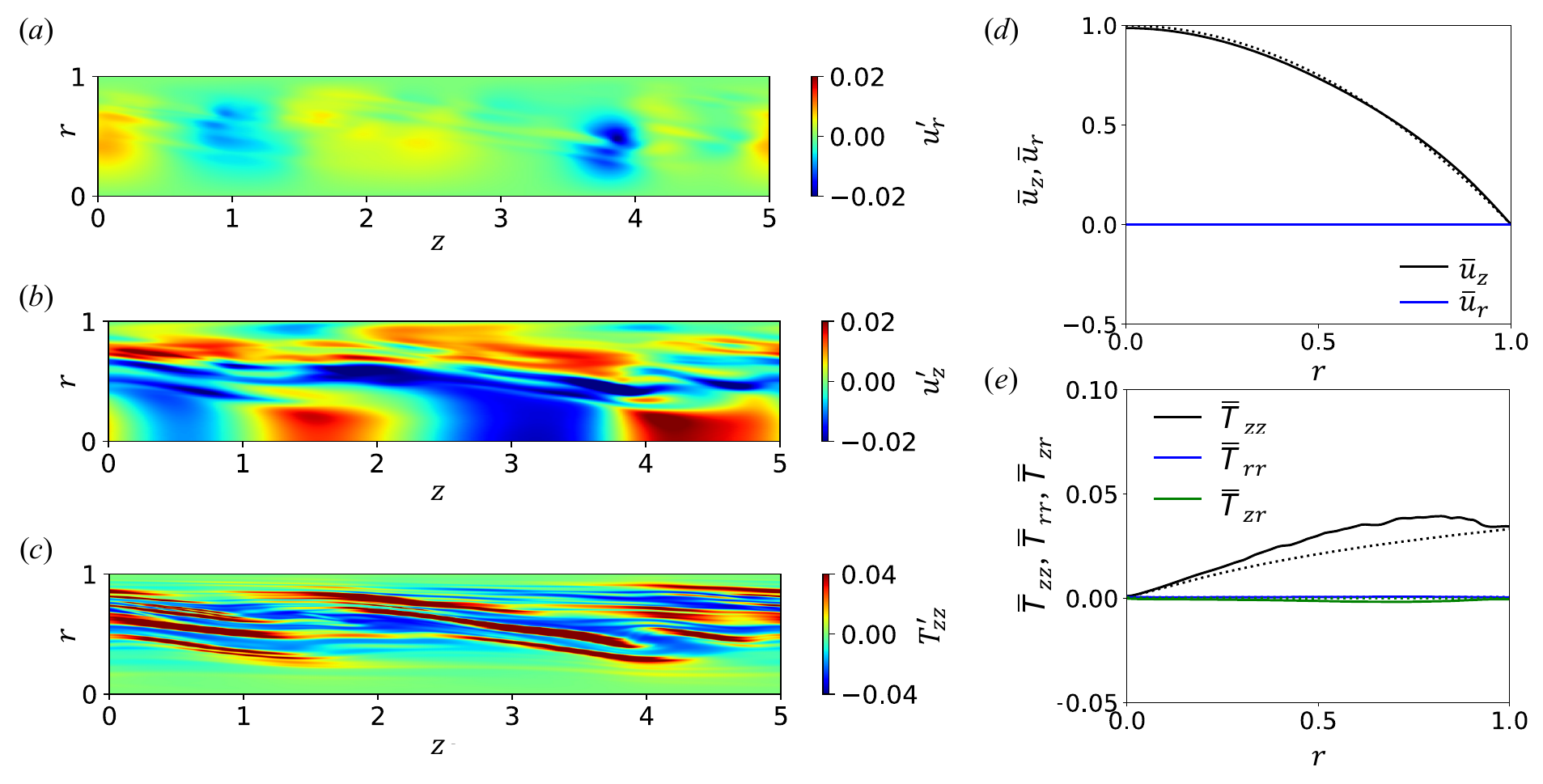}
\caption{Instantaneous ($a$) radial velocity, ($b$) streamwise velocity, and ($c$) $zz$-component of stretch tensor of elastoinertial turbulence (EIT) in axisymmetric pipe flow at $\Rey=3000$, $\Wi=35$, and $L=5$. Means (averaged over $z$ and $t$) of ($d$) velocity components and ($e$) stretch components. Dotted lines ($d,e$) represent the laminar profiles at the same parameter.} 
\label{EIT_snapshots_means_pipe_Wi35_Re3000.pdf}
\end{figure}

\section{Results and discussion}\label{results}
We use VESPOD analysis to uncover the structures underlying the dynamics of EIT in axisymmetric pipe flow at several parameter regimes and domain lengths. First, we analyze dynamics at $\Rey=3000$, $\Wi=35$, and $L=5$ (Section \ref{result_sec1}). Then, we discuss dynamics of EIT in a longer domain ($L=10$) (Section \ref{result_sec2}) and the effect of $\Rey$ and $\Wi$ on EIT (Section \ref{result_sec3}). Finally, we discuss the relationship between EIT in the pipe and channel flows and their link to linear dynamics (Section \ref{result_sec4}).   
 
\subsection{EIT at $\Rey=3000$, $\Wi=35$, and $L=5$} \label{result_sec1}

Figure \ref{EIT_snapshots_means_pipe_Wi35_Re3000.pdf} depicts the instantaneous snapshots of state variables and their temporal means for the dynamics of EIT at $\Rey=3000$, $\Wi=35$, and $L=5$. The dynamics of EIT has well-defined means (figure \ref{EIT_snapshots_means_pipe_Wi35_Re3000.pdf}($d,e$)). Therefore, we report the perturbations of different state variables from their temporal means. The dynamics consists of weak velocity fluctuations, consistent with the nature of EIT (figure \ref{EIT_snapshots_means_pipe_Wi35_Re3000.pdf}($a,b$)). The radial velocity field ($u_r'$) has distinct large-scale dynamic patterns dominating the region between the pipe centerline and the wall (figure \ref{EIT_snapshots_means_pipe_Wi35_Re3000.pdf}($a$), movie 1). The streamwise velocity fluctuation ($u_z'$) has thin elongated streaks close to the wall; however, it contains large-scale structures in the vicinity of the pipe centerline (figure \ref{EIT_snapshots_means_pipe_Wi35_Re3000.pdf}($b$), movie 2). The polymeric stress field contains the formation of thin sheets of high polymeric stress in the most part of the pipe with slight clustering near the pipe wall (Appendix \ref{mesh_resolution_study}, movie 3). In the present paper, we use the scaled stretch field ($\mathsfbi{T}$) to represent polymeric stress field (figure \ref{EIT_snapshots_means_pipe_Wi35_Re3000.pdf}($c$)) and due to the appropriate scaling of the stretch field, which is based on its contribution to the mechanical energy, the velocity and stretch fields have similar magnitude (figure \ref{EIT_snapshots_means_pipe_Wi35_Re3000.pdf}($a-c$)). This observation provides a quantitative evidence that for elastoinertial turbulence both inertia and elasticity are equally important. The $zz$-component of the polymeric stress (and stretch) tensor dominates the stress field. Therefore, we report only $T_{zz}'$ in the present study. The stretch field ($T_{zz}'$) contains thin-inclined sheets of high stretch fluctuations situated well away from the pipe center $r=0$ (figure \ref{EIT_snapshots_means_pipe_Wi35_Re3000.pdf}($c$)). 

\begin{figure}
\centering
\includegraphics[width=\textwidth]{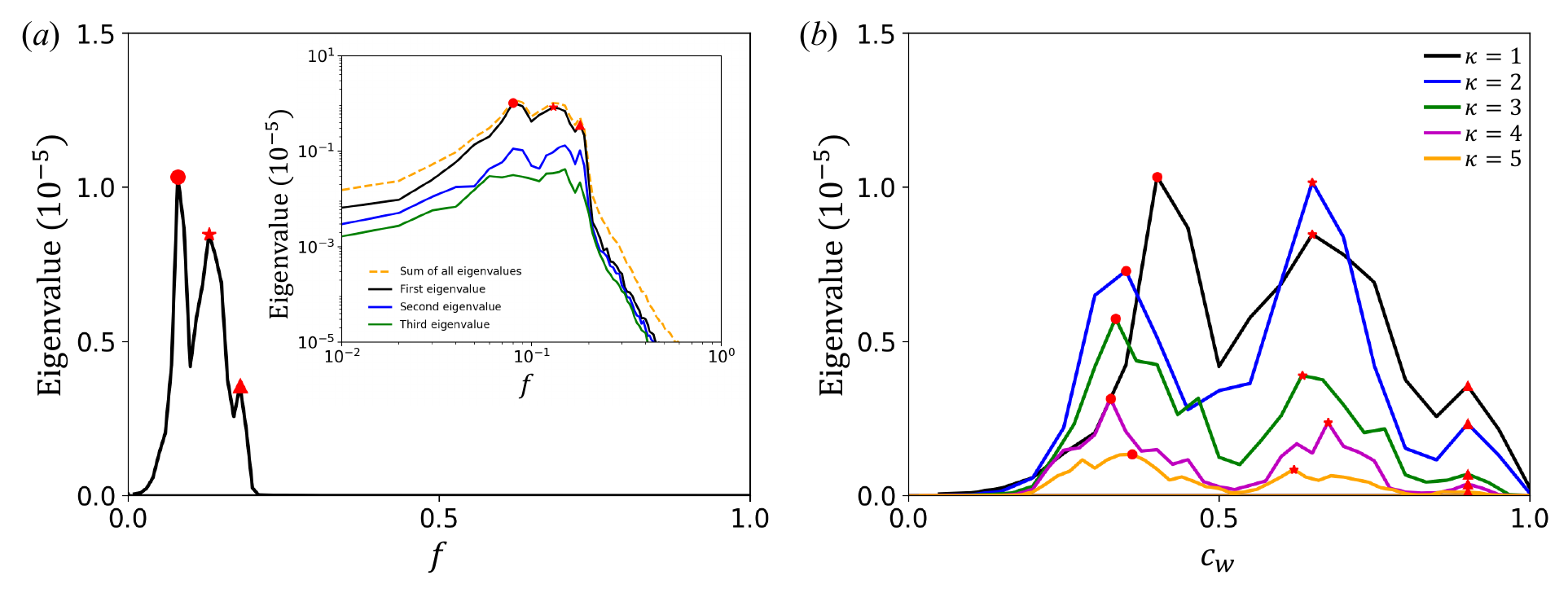}
\caption{($a$) Leading eigenvalue of the VESPOD energy spectrum of the structures having wavenumber $\kappa=1$ underlying EIT at $\Rey=3000$, $\Wi=35$, and $L=5$ (Inset: spectrum in log-log scale). ($b$) Leading eigenvalues of the VESPOD energy spectra for structures having different wavenumbers. Red symbols represent the peaks in the leading eigenvalues (Circle $\bullet$, star $\star$, and triangle $\blacktriangle$ are first, second, and third peaks, respectively).} 
\label{VESPOD_spectrum.pdf}
\end{figure}

\begin{figure}
\centering
\includegraphics[width=\textwidth]{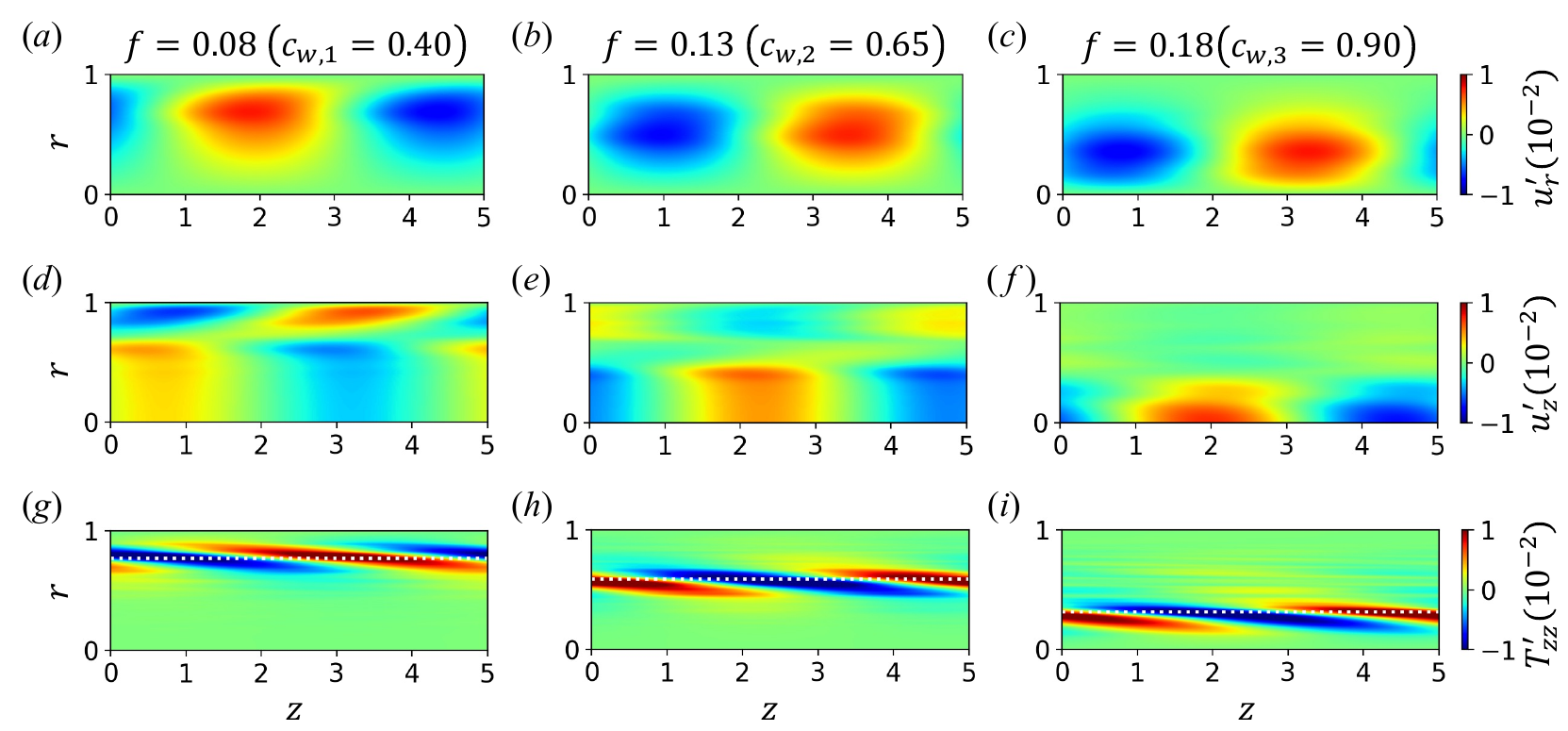}
\caption{VESPOD mode structures of ($a-c$) radial velocity ($u_r'$), ($d-f$) streamwise velocity ($u_z'$), and ($g-i$) $zz$-component of stretch tensor ($T_{zz}'$) corresponding to the peaks in the leading eigenvalue of VESPOD energy spectrum of $\kappa=1$. Locations of critical layers have been denoted with white dotted lines ($g-i$). Other parameters are $\Rey=3000$, $\Wi=35$, and $L=5$.} 
\label{mode_structure_mode_k1.pdf}
\end{figure}

The VESPOD energy spectrum of the structures having wavenumber $\kappa =1$ has been depicted in figure \ref{VESPOD_spectrum.pdf}($a$). The leading eigenvalue (integrated over all frequencies) contributes approximately $78 \%$ to the total variance (approximate mechanical energy), and there are three distinct peaks in the leading eigenvalue, suggesting that the mode structures corresponding to these peaks have a large contribution to the overall dynamics. In figure \ref{mode_structure_mode_k1.pdf}, we visualize the mode structures corresponding to different peaks in the energy spectrum. The radial velocity component has large-scale structures located away from the pipe centerline, which move from the location close to the pipe wall for the peak at low frequency to the vicinity of the pipe centerline for the peak at high frequency (figure \ref{mode_structure_mode_k1.pdf}($a-c$)). The streamwise velocity component has large-scale structures centered at the pipe centerline, which shrink in the radial direction as the frequency of peak increases; however, it has elongated patches of velocity fluctuations close to the wall (figure \ref{mode_structure_mode_k1.pdf}($d-f$)). The stretch field contains a layer having thin inclined sheets of high stretch fluctuations, and the location of this layer moves to the centerline of the pipe as the frequency increases (figure \ref{mode_structure_mode_k1.pdf}($g-i$)). These three structures are distinct traveling waves; their wave speeds $c_w=fL/\kappa$ can be given as $c_{w,1}$, $c_{w,2}$, and $c_{w,3}$ in the ascending order of their wave speed, and corresponding peaks in the energy spectrum as peak index 1, 2, and 3, respectively.

The concept of critical layer has been proven useful in the understanding of transitional flows, as for traveling fluctuations it provides the most favorable place to exchange energy between the base flow and fluctuations \citep{drazin1981hydrodynamic}. Therefore, we also explore the relationship between the critical layer of the traveling wave and the mode structures. As the mean velocity profile is close to the laminar parabolic profile (figure \ref{EIT_snapshots_means_pipe_Wi35_Re3000.pdf}($d$)), the location of the critical layer can be given as $r_c=\sqrt{1-c_w}$. The sheets of high polymeric stretch fluctuations in the VESPOD modes are localized around the location of the critical layer (figure \ref{mode_structure_mode_k1.pdf}($g-i$)). This observation is consistent with the structures of traveling waves underlying EIT in channel flow \citep{Shekar2019,Kumar2024}. 

\begin{figure}
\centering
\includegraphics[width=\textwidth]{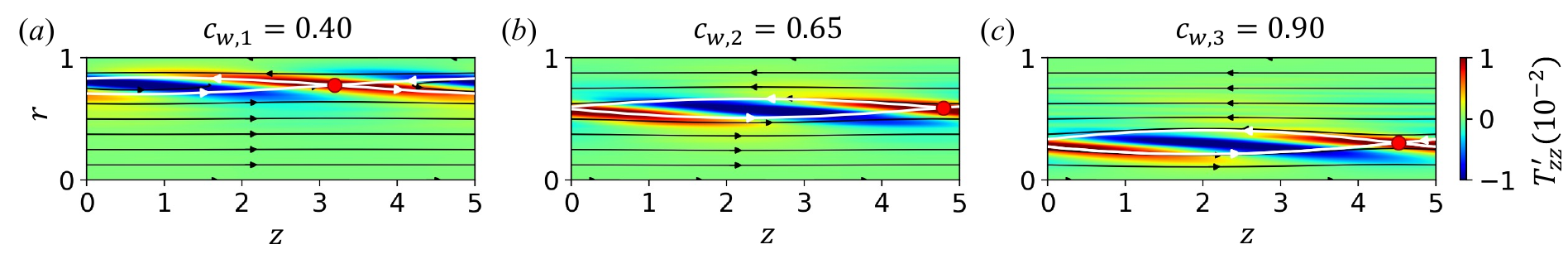}
\caption{\textcolor{black}{Streamlines of a velocity field consisting of the mean flow and a small component of the peak VESPOD mode at ($a$) $f=0.08 (c_{w,1}=0.40)$, ($b$) $f=0.13 (c_{w,2}=0.65)$, and ($c$) $f=0.18 (c_{w,3}=0.90)$ from figure \ref{VESPOD_spectrum.pdf}($a$), are shown in reference frames moving with their respective wave speeds, and are superimposed on the $zz$-component of the stretch tensor ($T_{zz}'$). The red circles denote hyperbolic stagnation points, and the streamlines passing through them are shown in white.}} 
\label{streamlines_stagnation_point.pdf}
\end{figure}

\textcolor{black}{To explore the mechanism of the emergence of sheets in the polymeric stress field, we plot the streamlines of velocity fields that are constructed from the mean flow plus a small contribution from the peak VESPOD mode at the peaks in figure \ref{VESPOD_spectrum.pdf}($a$), in the reference frame moving with their respective wave speeds, along with $T_{zz}'$ in the figure \ref{streamlines_stagnation_point.pdf}. There is a hyperbolic stagnation point in the velocity field (in the reference frame moving with the wave speed) at the critical layer of the traveling wave, where the flow is extension dominated. Polymeric chains get highly stretched as they approach and depart such stagnation point, leading to the formation of thin polymeric sheets at the critical layer. This critical layer mechanism is consistent with the critical layer mechanism of the origin of polymeric sheets in channel flow EIT \citep{Shekar2019}.}

\begin{figure}
\centering
\includegraphics[width=\textwidth]{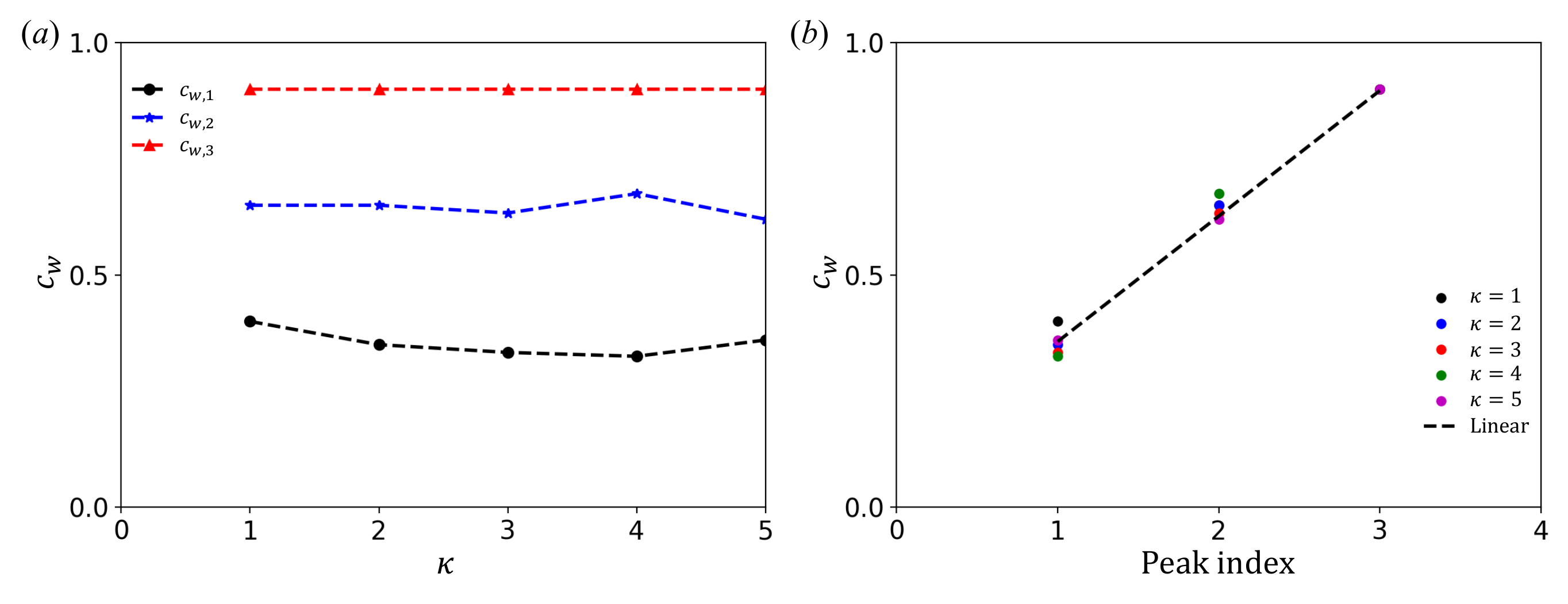}
\caption{Wave speeds of the VESPOD mode structures corresponding to the peaks in the energy spectra: ($a$) Wave speed vs wavenumber and ($b$) Wave speed vs peak index. Parameter values are $\Rey=3000$, $\Wi=35$, and $L=5$.} 
\label{wavespeed_vs_wavenumber.pdf}
\end{figure}

The VESPOD energy spectra of higher wavenumber structures have similar characteristics to those at $\kappa=1$. The leading eigenvalues contain most of the energy across the spectra of different wavenumber structures and have three distinct peaks. In figure \ref{VESPOD_spectrum.pdf}($b$), we visualize the leading eigenvalues of VESPOD corresponding to different $\kappa$ as a function of wave speed. For a given peak index, the wave speeds of the mode structures are nearly independent of wavenumber, implying that the higher wavenumber structures ($\kappa \geq 2$) are simple harmonics of the fundamental wavenumber structure ($\kappa =1$). This is also evident from figure \ref{wavespeed_vs_wavenumber.pdf}. The traveling waves corresponding to each peak index belong to the respective unique family of traveling waves having unique wave speed (figure \ref{wavespeed_vs_wavenumber.pdf}($a$)), and the wave speed of the family of traveling waves increases almost linearly with the peak index (figure \ref{wavespeed_vs_wavenumber.pdf}($b$)). 

\begin{figure}
\centering
\includegraphics[width=\textwidth]{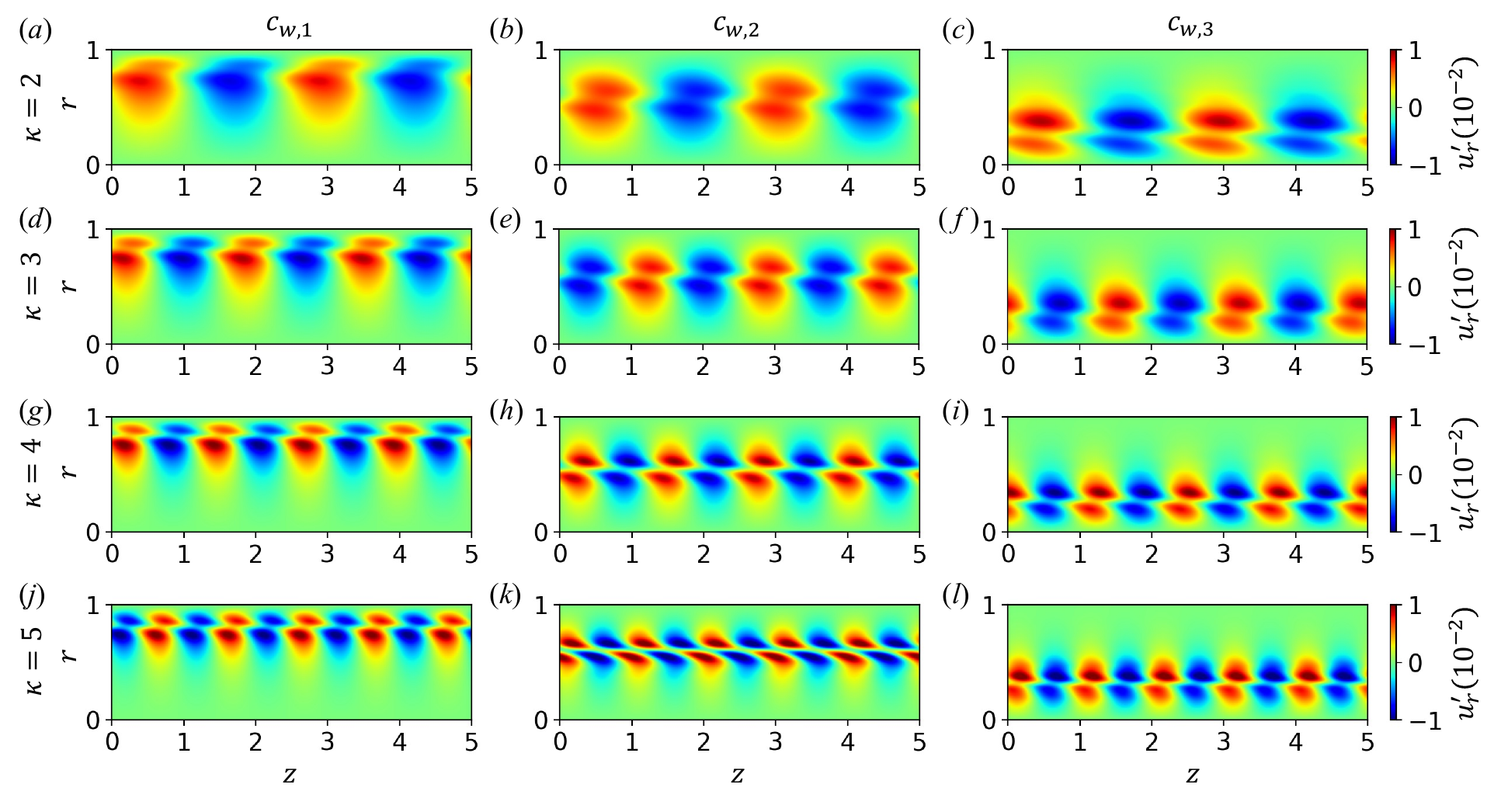}
\caption{VESPOD mode structures of $u_r'$ corresponding to different peaks in the leading eigenvalue of energy spectrum of wavenumber ($a-c$) $\kappa=2$, ($d-f$) $\kappa=3$, ($g-i$) $\kappa=4$, and ($j-l$) $\kappa=5$. Other parameters are $\Rey=3000$, $\Wi=35$, and $L=5$.} 
\label{mode_structure_uy_mode_k2to5.pdf}
\end{figure}

\begin{figure}
\centering
\includegraphics[width=\textwidth]{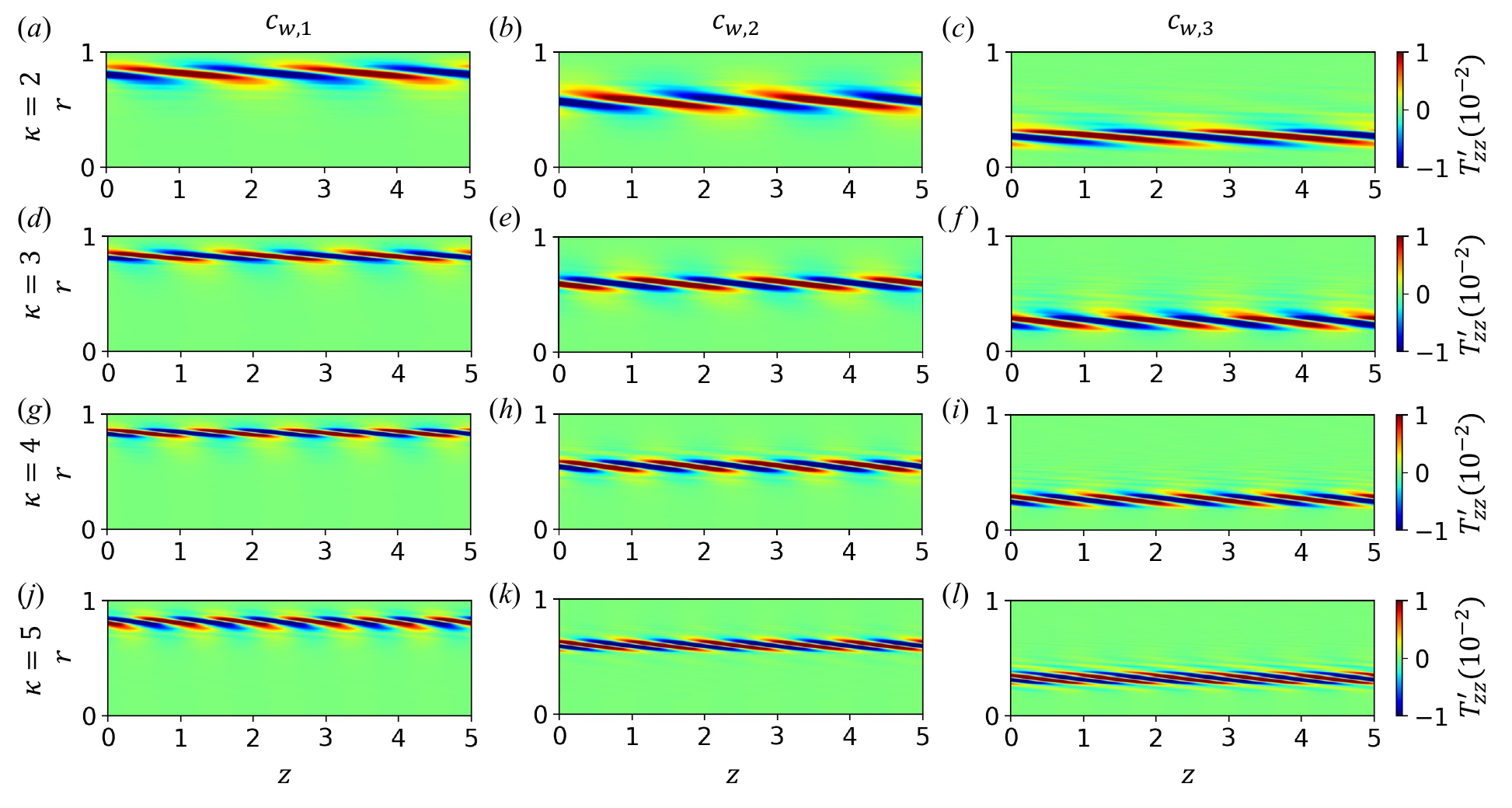}
\caption{VESPOD mode structures of $T_{zz}'$ corresponding to different peaks in the leading eigenvalue of energy spectrum of wavenumber ($a-c$) $\kappa=2$, ($d-f$) $\kappa=3$, ($g-i$) $\kappa=4$, and ($j-l$) $\kappa=5$. Other parameters are $\Rey=3000$, $\Wi=35$, and $L=5$. These structures correspond to the velocity structures shown in figure \ref{mode_structure_uy_mode_k2to5.pdf}.} 
\label{mode_structure_Txx_mode_k2to5.pdf}
\end{figure}

In addition, we depict $u_r'$ and $T_{zz}'$ of mode structures corresponding to the three distinct peaks in the VESPOD energy spectra of higher wavenumber structures ($\kappa \geq 2$) in figures \ref{mode_structure_uy_mode_k2to5.pdf} and \ref{mode_structure_Txx_mode_k2to5.pdf}, respectively. The radial velocity component has large-scale structures, and these structures have a partition line oriented in the streamwise direction that divides the structures in two parts (figure \ref{mode_structure_uy_mode_k2to5.pdf}). For each wavenumber, the location of the partition line moves to the centerline of the pipe as the wave speed increases. However, for a given wave speed (i.e., peak index), the location of the partition line is insensitive to the wavenumber. The stretch field is characterized by the formation of a layer, again localized at the critical layer position, having thin inclined sheets of high stretch fluctuations (figure \ref{mode_structure_Txx_mode_k2to5.pdf}). For a given wave speed, the locations of this layer for different wavenumber structures coincide with each other. However, they move to the pipe centerline as the wave speed increases for each wavenumber structure. Further, the location of partition line of the structure in radial velocity coincides with the critical layer, and hence the polymeric sheets.

The discussion so far demonstrates that the dynamics of EIT in pipe flow at $\Rey=3000$, $\Wi=35$, and $L=5$ is dominated by three distinct families of traveling waves, \textcolor{black}{where higher wavenumber structures ($\kappa \geq 2$) are the simple harmonics of the respective fundamental waves ($\kappa=1$).} 
There also exist some relationships among different families of traveling waves. For example, the sheets of high polymeric stretch fluctuations of $c_{w,1}$ are located in the vicinity of the wall of the pipe and the polymeric stretch sheets of $c_{w,2}$ are confined in the region bounded by the pipe centerline and polymeric stretch sheets of $c_{w,1}$. Similarly, the sheets of $c_{w,3}$ are bounded by the pipe centerline and the sheets of $c_{w,2}$. Thus, the polymeric sheets of different families of traveling waves exhibit a nested arrangement, where the polymeric sheets of faster waves are confined by the polymeric sheets of immediately slower waves. A similar nesting relationship among the dominant traveling waves underlying EIT in channel flow has been observed \citep{Kumar2024}. 

\subsection{EIT at $\Rey=3000$, $\Wi=35$, and $L=10$} \label{result_sec2}

\begin{figure}
\centering
\includegraphics[width=0.9\textwidth]{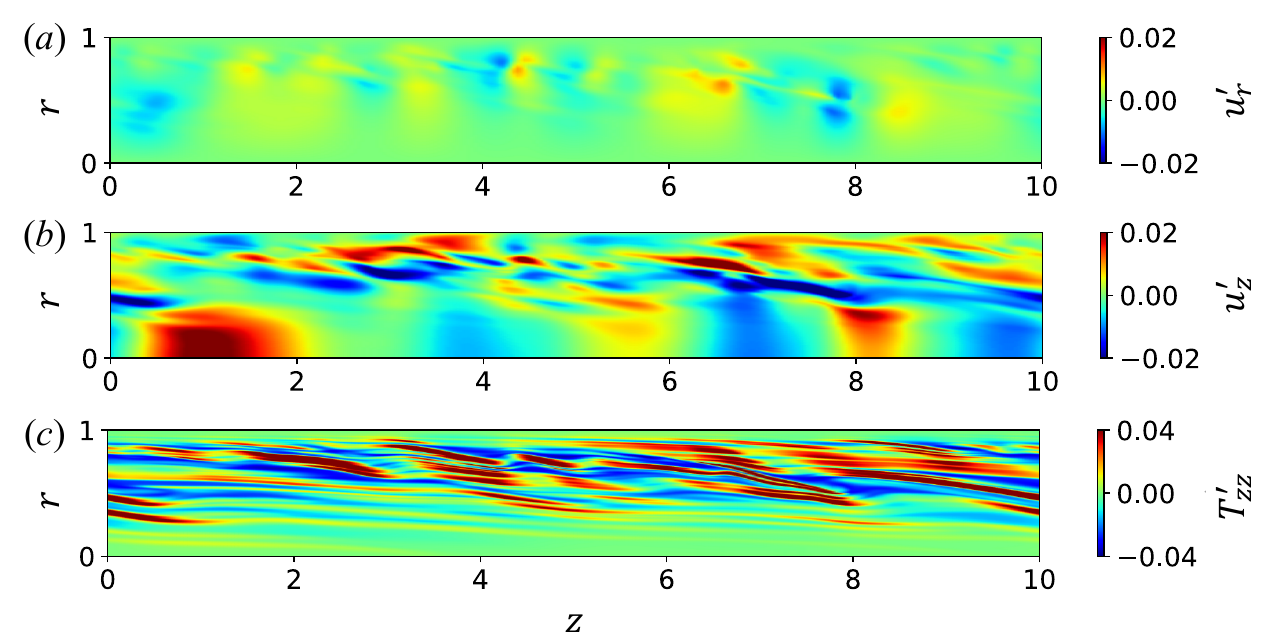}
\caption{\textcolor{black}{Snapshots of ($a$) radial velocity, ($b$) streamwise velocity, and ($c$) $zz$-component of stretch tensor of elastoinertial turbulence in axisymmetric pipe flow at $\Rey=3000, \Wi=35$, and $L=10$.}} 
\label{EIT_snapshots_means_pipe_Wi35_Re3000_L10.pdf}
\end{figure}

\begin{figure}
\centering
\includegraphics[width=\textwidth]{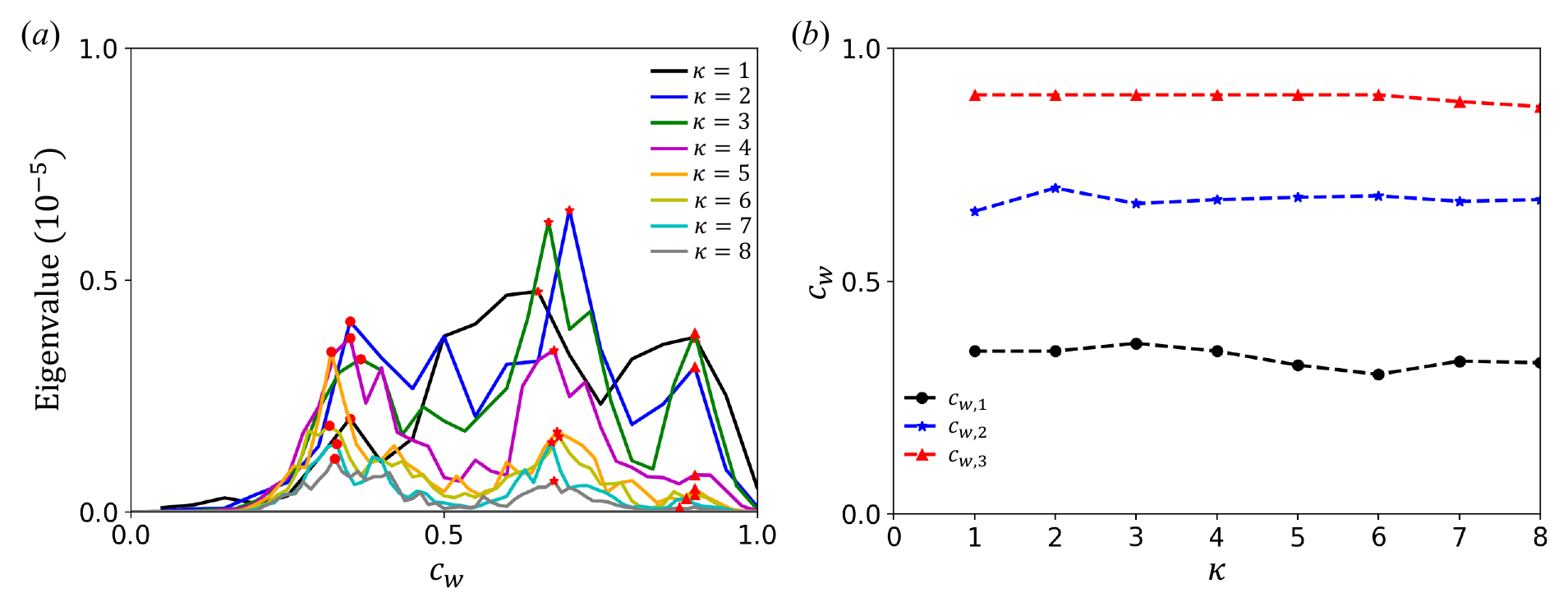}
\caption{($a$) Leading eigenvalues of the VESPOD energy spectra of different wavenumber structures at $\Rey=3000$, $\Wi=35$, and $L=10$. ($b$) Wave speeds of the traveling waves corresponding to the peaks in the leading eigenvalues of VESPOD energy spectra.} 
\label{eigenmode_vs_wave_speed_L10.pdf}
\end{figure}

\begin{figure}
\centering
\includegraphics[width=\textwidth]{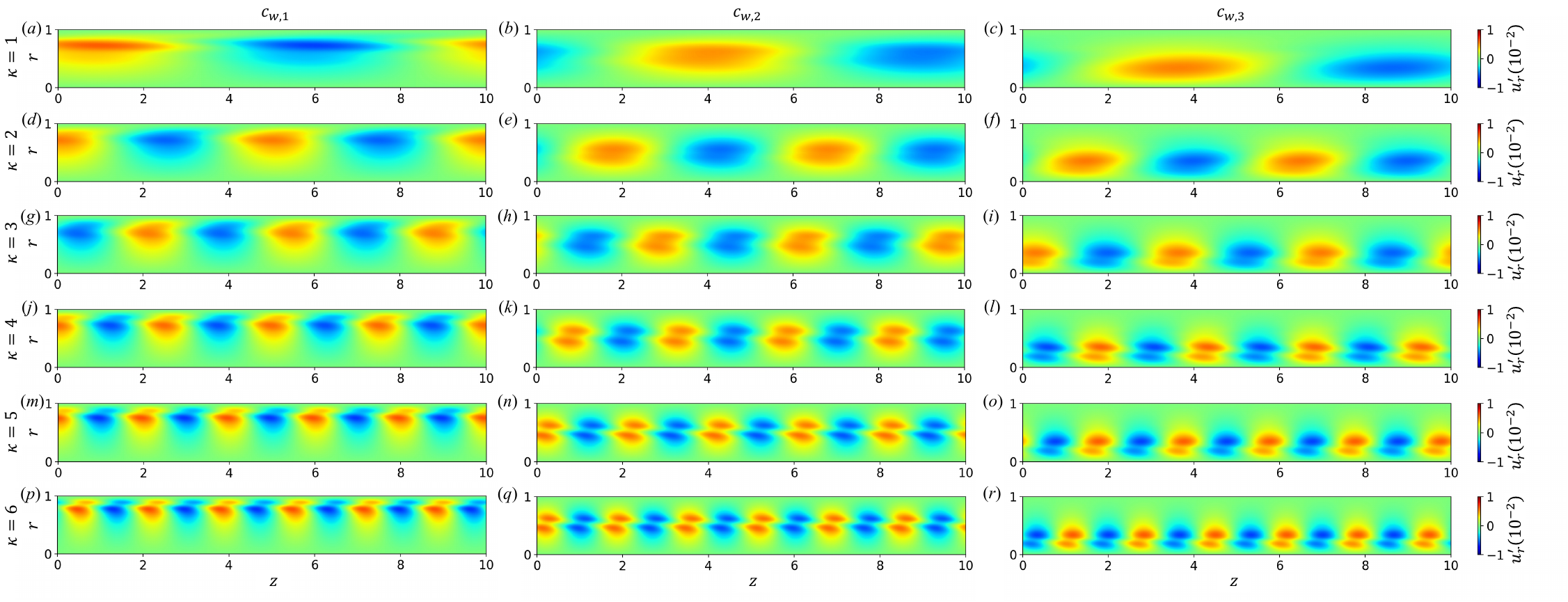}
\caption{VESPOD mode structures of $u_r^{\prime}$ corresponding to different peaks in the leading eigenvalue of energy spectrum of different wave number structures in the pipe of length $L=10$ ($\Rey=3000$ and $\Wi=35$).} 
\label{mode_structure_ur_mode_k1to6_L10.pdf}
\end{figure}

To explore the effects of pipe length on EIT dynamics, we performed a simulation in a domain twice the length of pipe discussed in Subsection \ref{result_sec1}, but at the same $\Rey$ and $\Wi$. \textcolor{black}{The velocity and stretch fields corresponding to EIT in the longer pipe are shown in figure \ref{EIT_snapshots_means_pipe_Wi35_Re3000_L10.pdf}. We analyze VESPOD spectra to identify the dominant coherent structures underlying it and how pipe length affects them.} The leading eigenvalue of the VESPOD spectrum of each wavenumber structure contains a dominant contribution to the total mechanical energy and has three distinct peaks similar to the shorter pipe. We visualize the leading eigenvalue of different wavenumber structures as a function of wave speed and find that the wave speeds corresponding to peaks in the leading eigenvalues are nearly independent of the wavenumber of structures (figure \ref{eigenmode_vs_wave_speed_L10.pdf}($a$)). This shows that traveling waves with high wavenumber ($\kappa \geq 2$) are harmonics of the fundamental wave ($\kappa=1$). This is also evident from figure \ref{eigenmode_vs_wave_speed_L10.pdf}($b$), where we see three distinct families of traveling waves originating from the simple harmonic excitation of respective fundamental waves. We have also visualized the structures of different traveling waves in figure \ref{mode_structure_ur_mode_k1to6_L10.pdf}. We would like to emphasize that the traveling waves having an even wavenumber ($\kappa=2,4, \ldots$) in the pipe of $L=10$ are the same traveling waves discovered in the pipe of $L=5$. However, the traveling waves with an odd wavenumber ($\kappa=1,3,5, \ldots$) in the pipe of $L=10$ are new and do not exist in the pipe of $L=5$ because they do not satisfy the periodicity in the shorter domain. We also note that the energy associated with $\kappa=2$ structures is higher than that of $\kappa = 1$ structures (at least for the first two families of traveling waves), suggesting that the structures having wavelength $5$ units still dominate even in the pipe of $L=10$ (figure \ref{eigenmode_vs_wave_speed_L10.pdf}($a$)). This indicates that the pipe of $L=5$ is long enough to capture the dominant structures.

\subsection{Effect of Re and Wi} \label{result_sec3}

\begin{figure}
\centering
\includegraphics[width=\textwidth]{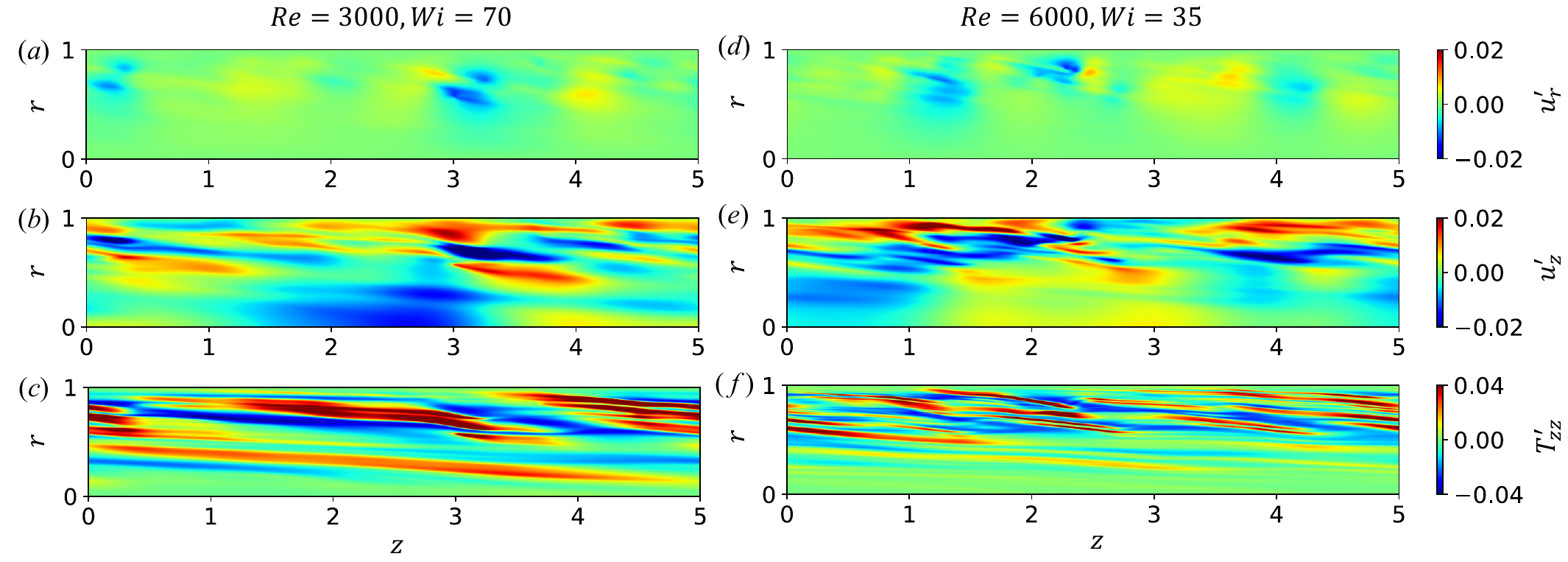}
\caption{\textcolor{black}{Snapshots of ($a,d$) radial velocity, ($b,e$) streamwise velocity, and ($c,f$) $zz$-component of stretch tensor of elastoinertial turbulence in axisymmetric pipe flow at ($a-c$) $\Rey=3000, \Wi=70$ and ($d-f$)  $\Rey=6000, \Wi=35$.}} 
\label{EIT_snapshots_means_pipe_Wi70_Re6000_5x1.pdf}
\end{figure}

\begin{figure}
\centering
\includegraphics[width=\textwidth]{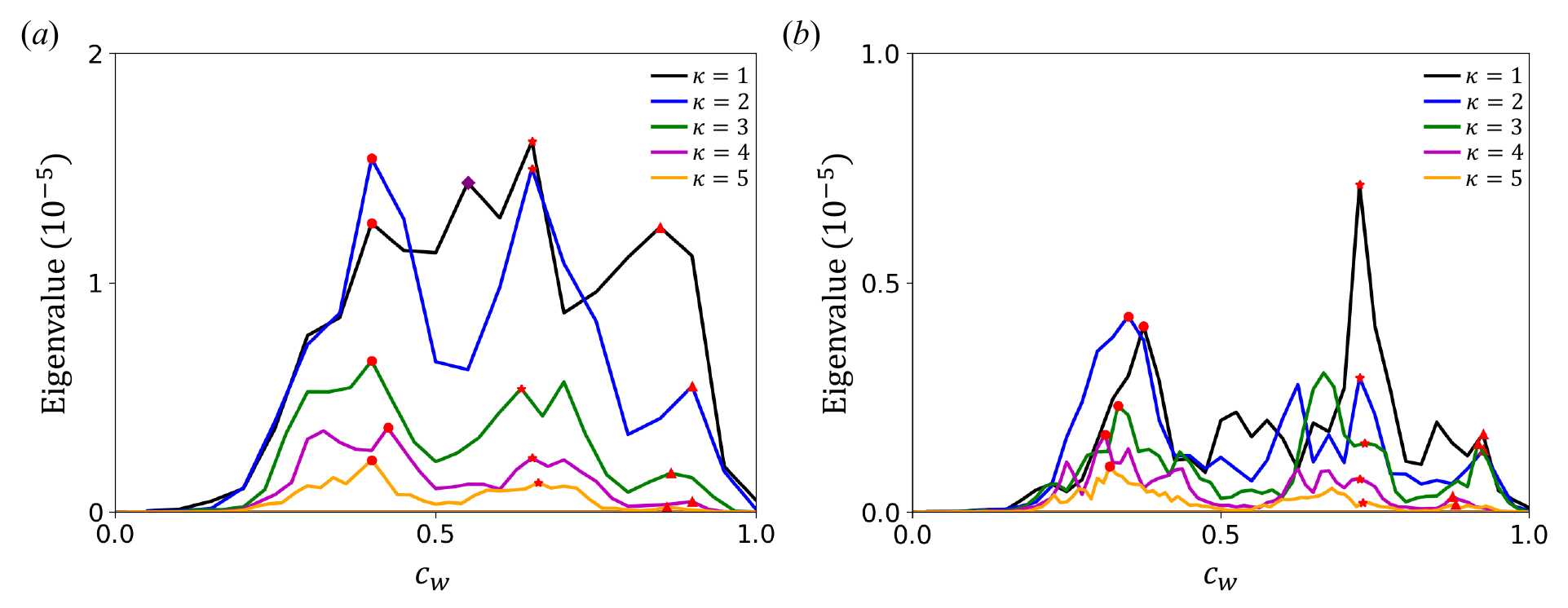}
\caption{Leading eigenvalues of VESPOD energy spectra for different wavenumber structures at ($a$) $\Rey=3000, \Wi=70$ and ($b$)  $\Rey=6000, \Wi=35$. \textcolor{black}{Red symbols represent harmonic peaks, while purple diamond symbol ($\blacklozenge$) represents a non-harmonic peak.}} 
\label{eigenmode_vs_wave_speed_Wi70_Re6000.pdf}
\end{figure}

\begin{figure}
\centering
\includegraphics[width=\textwidth]{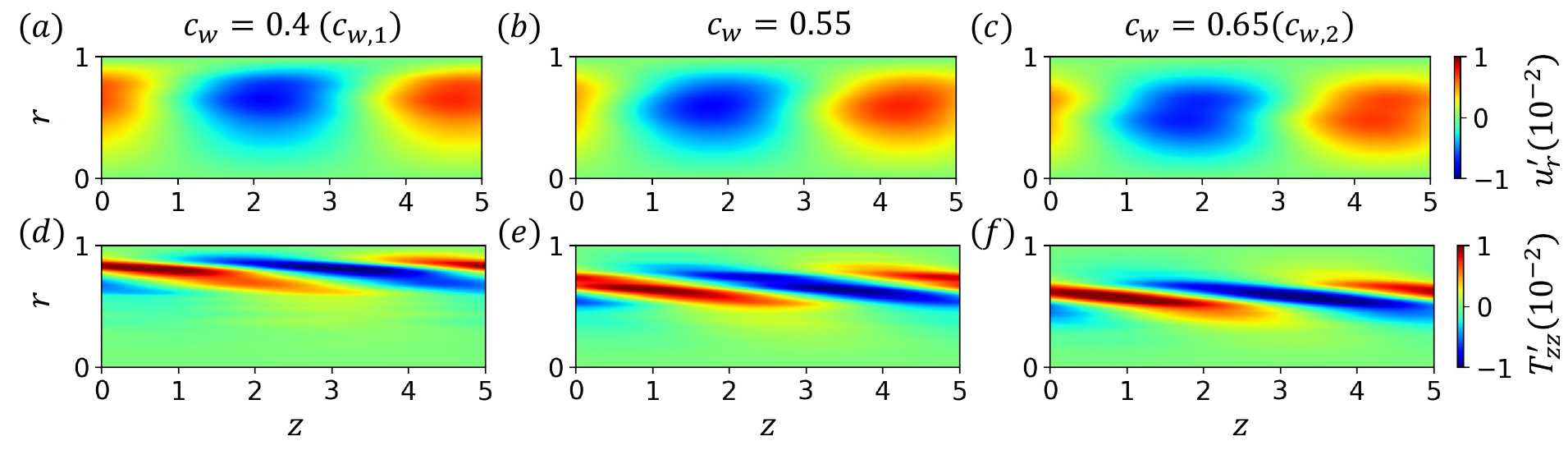}
\caption{VESPOD mode structures of ($a-c$) $u_r^{\prime}$ and ($d-f$) $T_{zz}^{\prime}$ corresponding to the first three peaks in the leading eigenvalue of $\kappa=1$ at $\Rey=3000$ and $\Wi=70$. The middle column corresponds to the structures of the non-harmonic peak (purple diamond ($\blacklozenge$) in figure \ref{eigenmode_vs_wave_speed_Wi70_Re6000.pdf}($a$)), however first and third columns represent harmonic peaks.} 
\label{mode_structure_Re3k_Wi70_k1.pdf}
\end{figure}

To explore the effects of $\Wi$ and $\Rey$ on the dynamics of EIT, we analyze EIT at ($\Rey=3000, \Wi=70, L=5$) (figure \ref{EIT_snapshots_means_pipe_Wi70_Re6000_5x1.pdf}($a-c$)) and ($\Rey=6000, \Wi=35, L=5$) (figure \ref{EIT_snapshots_means_pipe_Wi70_Re6000_5x1.pdf}($d-f$)). The VESPOD energy spectrum at ($\Rey=3000, \Wi=70$) exhibits characteristics similar to those of the spectrum at ($\Rey=3000, \Wi=35$) (figures \ref{eigenmode_vs_wave_speed_Wi70_Re6000.pdf}($a$) and \ref{VESPOD_spectrum.pdf}($b$)). The three distinct harmonic families of traveling waves persist even at $\Wi=70$, evident from the three distinct peaks in the leading eigenvalues of the energy spectra of different wavenumber structures. For EIT at $\Wi=70$, the speeds of these three families of traveling waves are the same as the waves at $\Wi=35$, suggesting that $\Wi$ does not affect the speeds of different traveling waves at a fixed $\Rey$. At higher value of $\Wi$, along with three harmonic peaks in the leading eigenvalues of VESPOD energy spectra, we also observe additional distinct peaks for some wavenumber structures (figure \ref{eigenmode_vs_wave_speed_Wi70_Re6000.pdf}($a$), purple diamond symbol ($\blacklozenge$)). These are non-harmonic modes, having energy comparable to the harmonic modes. The mode structures corresponding to the peaks at $\Wi=70$ are similar to $\Wi=35$. Therefore, here we only include the structures corresponding to the first three peaks in the leading eigenvalue of $\kappa=1$ at $\Rey=3000$ and $\Wi=70$ (figure \ref{mode_structure_Re3k_Wi70_k1.pdf}). The mode structures shown in figure \ref{mode_structure_Re3k_Wi70_k1.pdf} ($a,d$) and figure \ref{mode_structure_Re3k_Wi70_k1.pdf} ($c,f$) correspond to the fundamental wave of the first ($c_{w,1}$) and the second ($c_{w,2}$) families of harmonic traveling waves. However, the structures in figure \ref{mode_structure_Re3k_Wi70_k1.pdf} ($b,e$) correspond to the non-harmonic peak in the leading eigenvalue of the $\kappa=1$ structure at $\Rey=3000$ and $\Wi=70$ (figure \ref{eigenmode_vs_wave_speed_Wi70_Re6000.pdf}($a$)). We do not see any fundamental difference between the mode structures of the harmonic and non-harmonic peaks. 

Further, we analyze dynamics at higher $\Rey$ ($\Rey=6000$), while keeping $\Wi=35$, same as in Subsection \ref{result_sec1} (figure \ref{eigenmode_vs_wave_speed_Wi70_Re6000.pdf}($b$)). Again, there exist three dominant families of harmonic traveling waves. However, the speeds of the traveling waves at $\Rey=6000$ are slightly different than those at $\Rey=3000$. For example, the speed of the second family of traveling waves ($c_{w,2}$) at $\Rey=6000$ is higher than that at $\Rey=3000$. At larger $\Rey$, non-harmonic modes become increasingly important. \textcolor{black}{As $\Wi$ or $\Rey$ increases, the EIT dynamics becomes increasingly broadband and temporally complex, resulting in less distinct spectral peaks and weaker coherent traveling-wave signatures. Accordingly, additional peaks emerge alongside the harmonic peaks at higher $\Rey$ and $\Wi$, indicating increasingly irregular nonlinear dynamics.}

\subsection{Relation to channel flow EIT and linear dynamics}\label{result_sec4}
The dynamics of EIT in channel and pipe flows share several similarities, but also exhibit important differences. Even though both dynamics have similar first glance features such as the large-scale structures in wall-normal velocity and the formation of thin inclined sheets of high stress in the polymeric stress field, the dominant coherent structures underlying them are distinct. EIT in channel flow is dominated by a family of non-harmonic traveling waves, whereas in pipe flow it is dominated by the three distinct families of harmonic traveling waves. The wall-normal velocity component of these traveling waves exhibits large-scale structures centered at the centerline in channel flow, whereas they are located away from the centerline in the pipe flow. The streamwise velocity component of these waves exhibits regular structures in the vicinity of the centerline in pipe flow, whereas in channel flow such distinct structures are located away from the centerline. However, the polymeric stress field in both channel and pipe flows contains sheets of high stress fluctuations at the critical layers of the traveling waves. Further, the polymeric sheets of faster traveling waves in both channel and pipe flows are confined by the polymeric sheets of their immediately slower traveling waves, leading to a nested arrangement of polymeric sheets in both dynamics. 

\textcolor{black}{The resolvent analysis, which is based on the response of linearized governing equations to an external forcing, provides the modes that optimally describe the linear growth. It has been demonstrated that the SPOD modes are the same as the resolvent modes for uncorrelated resolvent-mode expansion coefficients, which is the case for the white-noise forcing of linear dynamics \citep{Towne2018}. The resolvent analysis for viscoelastic pipe flow considering the white noise forcing has found that the most strongly amplified resolvent mode exhibits thin tilted sheets like fluctuations in the polymeric stress field at the critical layer \citep{Zhang2021}, which closely resembles the structure of the most dominant traveling wave discovered here using the VESPOD analysis, suggesting a possible connection of the dominant coherent structures underlying EIT in pipe flow  with its linear dynamics. For channel flow, it has already been demonstrated that the Tollmien–Schlichting (TS) mode resulting from the linear stability analysis closely resembles the most strongly amplified resolvent mode \citep{Shekar2019} and the most dominant SPOD mode underlying EIT \citep{Kumar2024}.}

\textcolor{black}{The differences between the velocity fields of the dominant traveling waves underlying EIT in channel and pipe flows might be resulting from the differences in their underlying linear problem. The linear instability of a wall mode leads to the emergence of the TS wave in channel flow and the dominant traveling wave underlying EIT in channel flow resembles TS wave structure, whereas no analogous wall mode linear instability arises in pipe flow. However, the common critical-layer mechanism responsible for the formation of thin sheets in the polymeric stress field can explain the similarities observed in the structures of the polymeric stress field of the dominant traveling waves underlying EIT in both channel and pipe flows.}

\section{Conclusions}\label{conclusion}
In the present study, we reveal the dominant well-defined structures underlying the dynamics of elastoinertial turbulence (EIT) in axisymmetric pipe flow using a viscoelastic variant of spectral proper orthogonal decomposition (VESPOD). The VESPOD enables us to perform modal decomposition of velocity and polymeric stress fields together and identify coherently evolving mode structures that contribute the most to the total mechanical energy of the dynamics of EIT. The leading eigenvalue of the VESPOD spectrum contains most of the mechanical energy and hence dominates its dynamics. The leading eigenvalue of the VESPOD energy spectrum of each wavenumber ($\kappa$) structure exhibits three distinct peaks. Further analysis of the mode structures corresponding to these peaks reveals three distinct families of well-defined traveling waves, which are characterized by their wave speed, dominate the dynamics of elastoinertial turbulence in axisymmetric pipe flow. The higher wavenumber structures ($\kappa \geq 2$) of each family of traveling wave are harmonics of their respective fundamental waves ($\kappa=1$). The radial velocity component of these traveling waves contains large-scale structures spanning the radial direction, however the polymeric stress field contains thin-sheets of high stress fluctuations localized at the critical layers of the traveling waves. \textcolor{black}{The critical layers contain hyperbolic stagnation points (in the reference frame moving with the wave speeds), which lead to the emergence of thin polymeric sheets. The substantial portion of fluctuations in EIT travel downstream with the three wave speeds as the form of highly coherent traveling wave structures and the radial positions corresponding to their critical layers are the preferred locations for the formation of thin sheets in the polymeric stress field.} Further, the polymeric sheets of a given traveling wave confine the polymeric sheets of immediately faster wave, leading to a nested arrangement of the polymeric sheets. At larger values of the Reynolds numbers and Weissenberg numbers, the EIT dynamics becomes more chaotic, leading to the emerge of non-harmonic traveling waves, along with the three families of harmonic traveling waves, having energy content comparable to the harmonic waves.

\section*{Acknowledgments}
This research was supported under Office of Naval Research grant N00014-18-1-2865 (Vannevar Bush Faculty Fellowship) and National Science Foundation grant CBET-2437151.

\section*{Declaration of Interests}
The authors report no conflict of interest. 
 
\appendix

\section{Validation of numerical tool}\label{code_validation}
\begin{figure}
\centering
\includegraphics[width=0.5\textwidth]{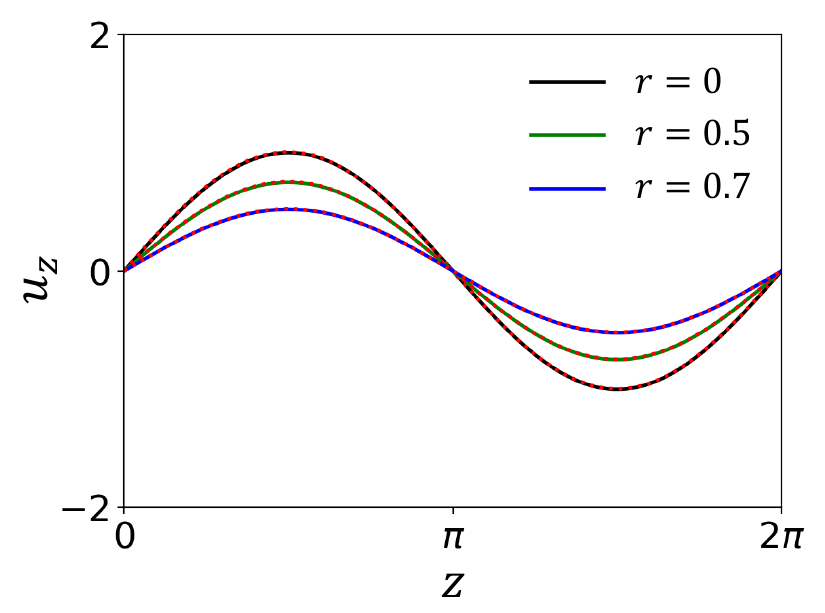}
\caption{Exact (solid lines) and computed (dotted lines) profiles of $u_z$ for a manufactured solution of viscoelastic flow through axisymmetric pipe at $\Rey=3000$, $\Wi=35$, $\beta=0.97$, and $b=6400$ (mesh resolution $256 \times 512$).} 
\label{code_validation_axisymmetric_pipe_flow.pdf}
\end{figure}

We use the method of the manufactured solution for the validation of the numerical tool. We assume a solution having velocity and conformation tensor fields as: 
\begin{equation}\label{manufactured_sol_u}
u_z=\left(1-r^2\right)sin(z), \qquad u_r=0 
\end{equation}
and 
\begin{equation}\label{manufactured_sol_alpha}
\boldsymbol{\alpha}=(1+sin(z))\boldsymbol{I}.
\end{equation}
Next, we analytically obtain forcing functions to the governing equations (\ref{colm} and \ref{fenep}) that satisfy the assumed solution (\ref{manufactured_sol_u} and \ref{manufactured_sol_alpha}). Using these forcing terms, we perform a numerical simulation and find a good agreement between the computed solution and the analytical solution (figure \ref{code_validation_axisymmetric_pipe_flow.pdf}).        

\section{Mesh resolution and $Sc$ dependency of solution}\label{mesh_resolution_study}

\begin{figure}
\centering
\includegraphics[width=.8\textwidth]{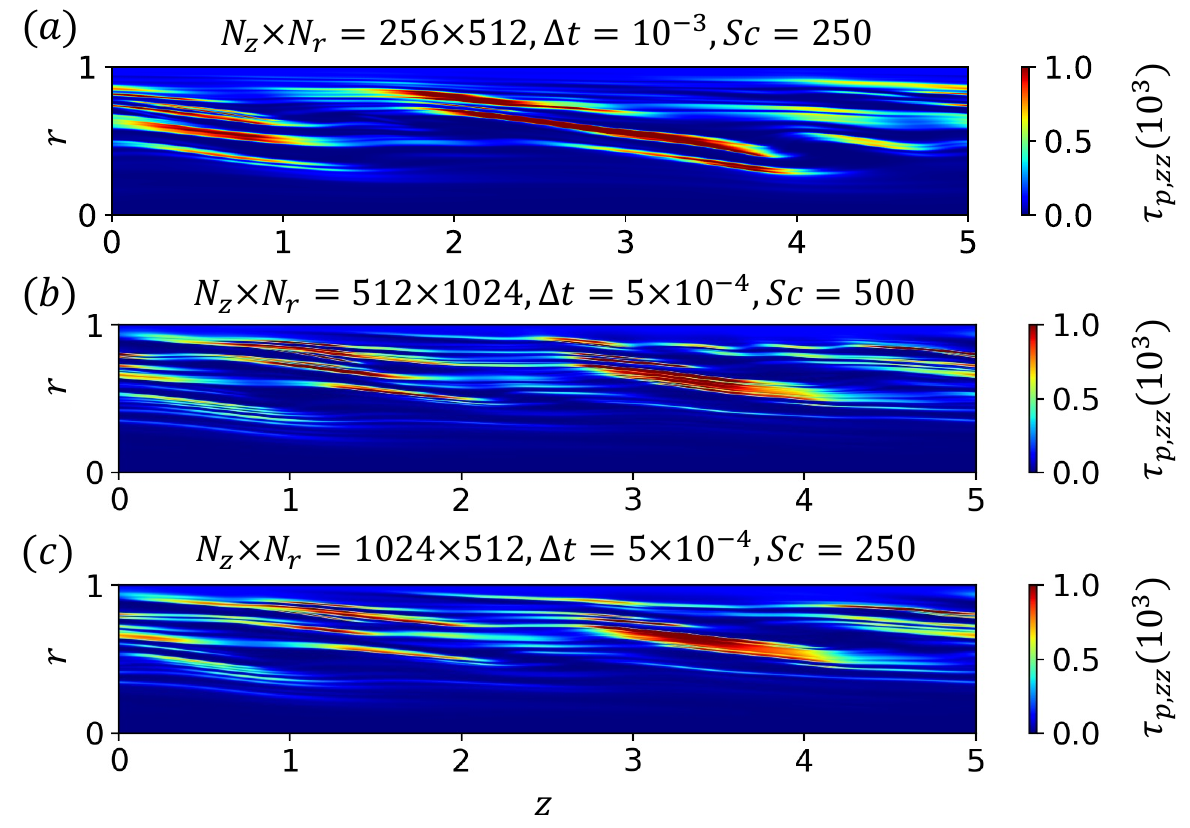}
\caption{\textcolor{black}{Snapshots of the $zz$-component of the polymeric stress field for EIT in an axisymmetric pipe flow obtained using different mesh resolutions and $Sc$ at the physical parameter values $\Rey=3000$, $\Wi=35$, and $L=5$.}} 
\label{tauzz_mesh_resolution_sc.pdf}
\end{figure}

\textcolor{black}{To illustrate the impact of mesh resolution and $Sc$ on EIT result, we depict the snapshots of the $zz$-component of the polymeric stress field (as this component is the most sensitive) obtained using different $Sc$ and the respective mesh resolution and time-step required to successfully evolve the EIT dynamics in axisymmetric pipe flow (figure \ref{tauzz_mesh_resolution_sc.pdf}). As the value of $Sc$ increases, polymeric stress diffusion becomes weaker and hence we see the emergence of even finer/sharper filaments oriented along the streamwise direction. The computation at higher $Sc$ becomes significantly expensive due to the requirement of higher mesh resolution and a smaller time-step to successfully evolve the dynamics. However, it does not fundamentally change the essential features of EIT and hence the central conclusion of the study. Therefore, in the present study, we use $Sc=250$, $N_z \times N_r=256 \times 512$, and $\Delta t=0.001$ as these simulation parameters provide a reasonable balance between accuracy and computational cost.}

\section{Probability Distribution Function (PDF) of tr$(\boldsymbol{\alpha})/b$ }\label{pdf_polymeric_stress}
\begin{figure}
\centering
\includegraphics[width=0.5\textwidth]{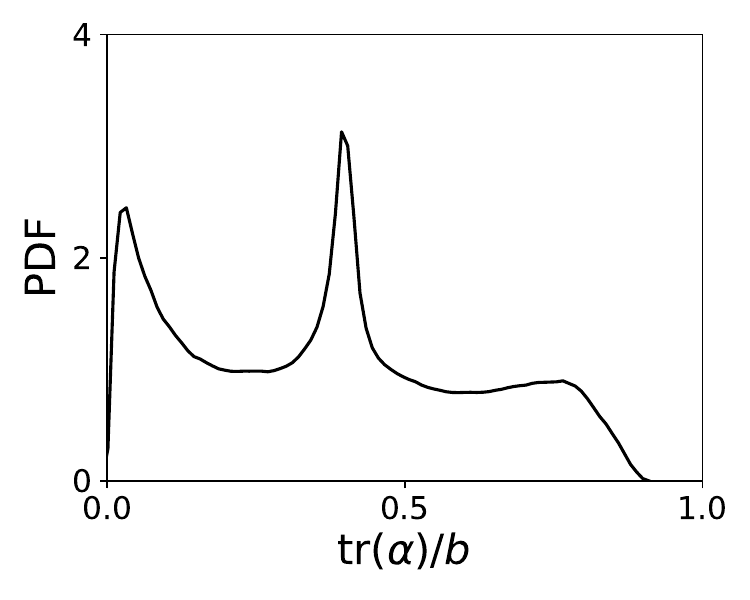}
\caption{\textcolor{black}{Probability Distribution Function (PDF) of tr$(\boldsymbol{\alpha})/b$ for polymeric stress field in EIT at $\Rey=3000$, $\Wi=35$, $\beta=0.97$, and $b=6400$.}} 
\label{pdf_trthetab_Re3000_Wi35_EIT_5x1_s3fm400.pdf}
\end{figure}

\textcolor{black}{To access the deviation of mechanical energy from the exact value due to the finite $b$ for the inner product used in the present study (Eq. \ref{mechanical_energy}), we estimate the PDF  of tr$(\boldsymbol{\alpha})/b$ for the ensemble of polymeric stress field in EIT. We define PDF as
\begin{equation}\label{pdf_tralpha_b}
PDF(\Tilde{p}) = \lim_{t\to\infty} \frac{\int_0^t \int_0^L \int_0^1 \delta (p(t,z,r)-\Tilde{p})rdrdzdt}{\int_0^t \int_0^L \int_0^1 rdrdzdt},
\end{equation}
where $p=$ tr$(\boldsymbol{\alpha})/b$, $\Tilde{p}$ is the value of $p$ at which PDF is evaluated, and $\delta$ is the Dirac delta function. Although the maximum value of tr$(\boldsymbol{\alpha})/b$ is $\approx 0.9$, the probability of a high value of tr$(\boldsymbol{\alpha})/b$ is small (figure \ref{pdf_trthetab_Re3000_Wi35_EIT_5x1_s3fm400.pdf}). In fact, the peaks in the PDF are around $\Tilde{p}=0$ and $\Tilde{p}=0.4$, suggesting the inner product used in the present study is a good representative of the mechanical energy. Regardless of whether the inner product used in present study exactly captures the mechanical energy, it is an appropriate and physically motivated inner product and norm.}

\section{VESPOD spectra based on different frequency resolution}\label{VESPOD_spectra_different_nfft}
\begin{figure}
\centering
\includegraphics[width=0.5\textwidth]{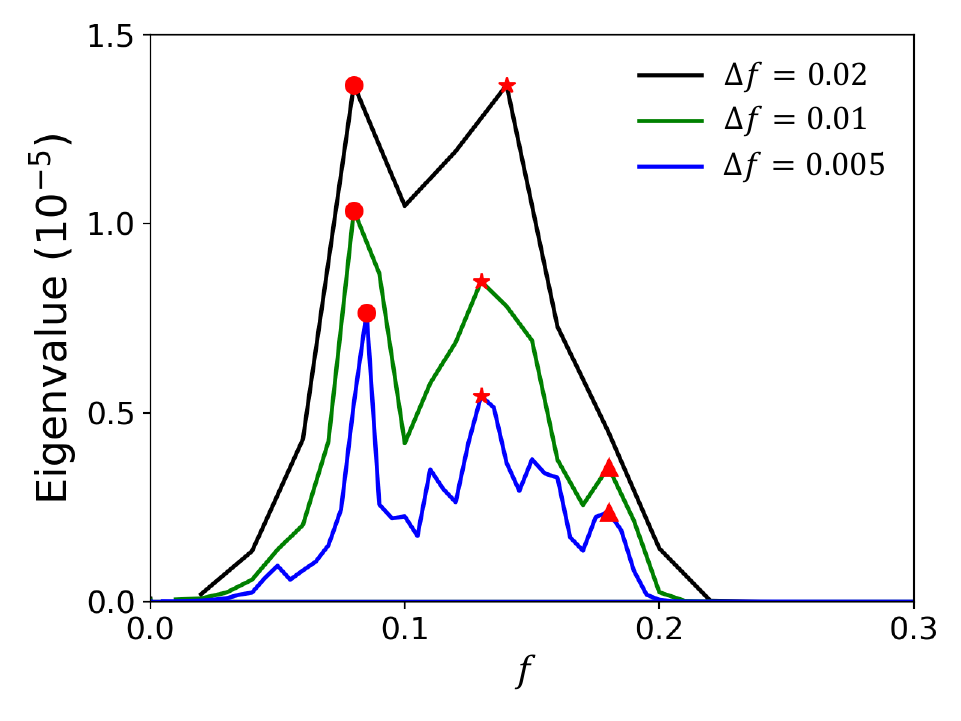}
\caption{\textcolor{black}{Leading eigenvalues of VESPOD spectra of $\kappa=1$ structures obtained using different frequency resolutions at $\Rey=3000$, $\Wi=35$, and $L=5$. Total 24000 snapshots were used for the frequency resolutions $\Delta f=0.02$ and $\Delta f=0.01$, whereas 36000 snapshots were used for $\Delta f=0.005$ to maintain statistical convergence.}} 
\label{VESPOD_Re3k_Wi35_nfft2k4k8k_snapshots24k24k41k.pdf}
\end{figure}

\textcolor{black}{For wavenumber $\kappa=1$ structures, the difference between the consecutive peak frequency is small, therefore, a higher frequency resolution is required to properly resolve the dominant peaks in the VESPOD spectra. We have plotted the leading VESPOD eigenvalue spectra for $\kappa=1$ structures obtained using different frequency resolutions (figure \ref{VESPOD_Re3k_Wi35_nfft2k4k8k_snapshots24k24k41k.pdf}). The frequency resolution $\Delta f=0.02$ is too coarse to resolve all the peaks associated with $\kappa=1$ structures, whereas $\Delta f=0.005$ requires larger dataset for statistical convergence and also generates some additional peaks resolving weak variability within broadband structures. Therefore, we choose $\Delta f=0.01$ for $\kappa = 1$ structures to balance frequency resolution and statistical convergence, enabling clearer identification of the dominant energetic structures.}

\bibliographystyle{jfm}
\bibliography{main-nested,turbulence-MDG-2402}

\end{document}